\def\llncs{0}
\def\mnotes{0}
\documentclass[11pt]{article}

\usepackage{amsfonts,amsmath,epsfig, graphicx, fullpage}
\usepackage{color,bm}

\def\colorson{0}

\usepackage{times}
\usepackage{color,xcolor}
\usepackage{amsfonts,amsmath}
\usepackage{url}
\usepackage{fullpage}

 \ifnum\llncs=0

\definecolor{orange}{RGB}{127,127,255}

\colorlet{red}{red!40!black}

\colorlet{orange}{blue!90!black}

\usepackage{amsthm}
\usepackage{latexsym,amssymb}
\usepackage{graphicx}

    \usepackage[letterpaper=true,pdftex,
    bookmarks,
    plainpages=false, 
 pdfpagelabels=true, 
 colorlinks=true,breaklinks=true,linkcolor=black, citecolor=orange,filecolor=black,menucolor=black,pagecolor=black,urlcolor=blue
    pdfauthor=Jalaj Upadhyay	
      ]{hyperref}{}

\fi 
\ifnum\mnotes=0
\newcommand{\mnote}[1]{}
\else
\newcounter{mynotes}
\newcommand{\mnote}[1]{\addtocounter{mynotes}{1}{{\bf !}}%
\marginpar{\scriptsize  {\arabic{mynotes}.\ {\sf \textcolor{red}{#1}}}}}
\fi

\ifnum\colorson=1
\newenvironment{todo}{\noindent
\sf \footnotesize \textcolor{blue}{To go here}:
\begin{CompactItemize}\color{blue}}
{\color{black}\end{CompactItemize}\rm \normalsize}
\else

\fi

\ifnum\llncs=0

\newenvironment{CompactItemize}{
  \vspace{-5pt}
 \begin{list}{$\bullet$}{%
      \setlength{\leftmargin}{12pt}%
      \setlength{\itemsep}{-2pt}
      }}
{\end{list}}

\makeatletter
\def\thm@space@setup{\thm@preskip=3pt
\thm@postskip=2pt}
\makeatother
\newtheoremstyle{newstyle}      
{} 
{} 
{\mdseries} 
{} 
{\bfseries} 
{.} 
{ } 
{} 

\theoremstyle{newstyle}

\newtheorem{theorem}{Theorem}[section]
\newtheorem{lemma}[theorem]{Lemma}

\newtheorem{prob}{Problem}

\newtheorem{definition}{Definition}[section]

\theoremstyle{definition}

\theoremstyle{remark}

\newtheorem{myremark}{Remark} 
\newenvironment{remark}{\begin{myremark}}{$\Box$\end{myremark}}

\newtheorem{myexample}{Example}

\else
\spnewtheorem{protocol}{Protocol}{\bfseries}{\rmfamily}
\spnewtheorem{algm}{Algorithm}{\bfseries}{\rmfamily}
\spnewtheorem{fact}{Fact}{\bfseries}{\rmfamily}
\spnewtheorem{myclaim}{Claim}{\bfseries}{\itshape}
\fi

\newcommand{\thmref}[1]{Theorem~\ref{thm:#1}}

\newcommand{\lemref}[1]{Lemma~\ref{lem:#1}}

\newcommand{\secref}[1]{Section~\ref{sec:#1}}

\newcommand{\appref}[1]{Appendix~\ref{app:#1}}

\newcommand{\figref}[1]{Figure~\ref{fig:#1}}

\newcommand{\eqnref}[1]{equation~(\ref{eq:#1})}

\newcommand{\comment}[1]{}
\newcommand{\ignore}[1]{}

\newcommand{\ket}[1]{|#1  \rangle}
\newcommand{\bra}[1]{\langle#1 |}
\newcommand{\tr}{\mbox{Tr}}

\newcommand{\brak}[1]{{\langle {#1} \rangle}}

\newcommand{\paren}[1]{\left( {#1} \right)}
\newcommand{\sparen}[1]{\left[ {#1} \right]}

\DeclareMathOperator{\poly}{poly}

\newcommand{\E}{\mathbb{E}}

\newcommand{\I}{\mathbb{I}}

\newcommand{\p}{\mathsf{Pr}}

\newcommand{\R}{\mathbb{R}}

\newcommand{\V}{\mathbb{V}}

\newcommand{\bX}{\textbf{X}}
\newcommand{\by}{\textbf{y}}
\newcommand{\bx}{\textbf{x}}

\newcommand{\cD}{\mathcal{D}}

\newcommand{\cK}{{\mathcal{K}}}

\newcommand{\cN}{\mathcal{N}}

\newcommand{\cQ}{\mathcal{Q}}

\newcommand{\cS}{\mathcal{S}}

\newcommand{\beq}{\begin{equation}}
\newcommand{\eeq}{\end{equation}}
\newcommand{\bml}{{\begin{multline}}}
\newcommand{\eml}{{\end{multline}}}

\usepackage[top=1in, bottom=1in, left=1in, right=1in]{geometry}

\usepackage{comment}

\includecomment{full}
\excludecomment{extended}

\makeatletter
\renewcommand{\paragraph}{%
 \@startsection{paragraph}{4}%
  {\z@}{1ex \@plus 1ex \@minus .2ex}{-1em}%
  {\normalfont\normalsize\bfseries}%
}
\makeatother

\usepackage{titlesec}

\begin{document}
\title{Differentially Private Linear Algebra in the Streaming Model}
\author{Jalaj Upadhyay \\
Center for Applied Cryptographic Research \\
David R. Cheriton School of Computer Science \\
University of Waterloo. \\
\small{\sf jalaj.upadhyay@uwaterloo.ca}}
\date{}
\maketitle

\newcommand{\cov}{\mathsf{COV}}
\newcommand{\PDF}{\mathsf{PDF}}
\newcommand{\range}{\mathsf{range}}
\newcommand{\Diag}{\mathsf{Diag}}

\newcommand{\bOmega}{\mathbf{\Omega}}
\newcommand{\bLambda}{\mathbf{\Lambda}}
\newcommand{\bSigma}{\mathbf{\Sigma}}
\newcommand{\bPhi}{\mathbf{\Phi}}
\newcommand{\bPsi}{\mathbf{\Psi}}
\newcommand{\bA}{\mathbf{A}}
\newcommand{\bD}{\mathbf{D}}
\newcommand{\bR}{\mathbf{R}}
\newcommand{\bY}{\mathbf{Y}}
\newcommand{\bU}{\mathbf{U}}
\newcommand{\bV}{\mathbf{V}}

\newcommand{\mult}{{\scshape Mat-Mult}}
\newcommand{\linear}{{\scshape Lin-Reg}}
\newcommand{\lra}{\mathsf{LRA}}
\newcommand{\low}{{\scshape LRA}}

\newcommand{\Comment}[2][red]{\marginpar{
   \fcolorbox{#1}{white}{\parbox{.8in}{
           \color{#1} \small \sf  #2 }}}}
\newcommand{\mcomment}[2]{\textcolor{red}{#1}\Comment{#2}}
\newcommand{\TODO}[2][\relax]{\noindent\mcomment{#2}{#1 TODO}}
\newcommand{\mytodo}[2]{\mcomment{ #1 }{#2 TODO}}
\newcommand{\ucomment}[3]{\par\noindent{\textcolor{#1}{/* #2 */}}\Comment[#1]{// #3}}

\newcommand{\Jcom}[1]{\ucomment{red}{#1}{Jalaj}}
\renewcommand{\Jcom}[1]{}

\maketitle

\begin{abstract}
Numerical linear algebra plays an important role in computer science. In this paper, we initiate the study of performing linear algebraic tasks while preserving privacy when the data is streamed online. Our main focus is the space requirement of the privacy-preserving data-structures. We give the first {\em sketch-based} algorithm for differential privacy. We give optimal, up to logarithmic factor, space data-structures that can compute low rank approximation, linear regression, and matrix multiplication,  while preserving differential privacy with better additive error bounds compared to the known results. Notably, we match the best known space bound in the non-private setting by Kane and Nelson (J. ACM, 61(1):4). 

Our mechanism for differentially private low-rank approximation {\em reuses} the random Gaussian matrix in a specific way to provide a single-pass mechanism. We prove that the resulting distribution also preserve differential privacy. This can be of independent interest. We do not make any assumptions, like singular value separation or normalized row assumption, as made in the earlier works. The mechanisms for matrix multiplication and linear regression can be seen as the private analogues of the known non-private algorithms. All  our mechanisms, in the form presented, can also be computed in the distributed setting. 
\end{abstract}

\pagebreak
\pagenumbering{arabic}

\section{Introduction} \label{sec:intro}
An $n \times d$ real-valued matrix is a natural structure for storing data about $n$ entities described by $d$ features. They arise in many contexts like information retrieval, data analysis, finance, scientific computation, genetics, and machine learning. In most of these applications, one is often required to do various linear algebraic tasks, like
low-rank matrix approximations ($\lra$), linear regression (\linear), and matrix multiplication (\mult).  
For example, $\lra$ is used in latent semantic indexing to speed-up the computation by computing a new representation for each document in the original collection, \linear \ is used in finance to analyze and quantify the systematic risk of an investment, and many matrix operations in scientific computation such as Gaussian elimination, $\mathsf{LU}$ decomposition, and the determinant or the inverse of a matrix can be  reduced to \mult. 


Let us consider some applications of these linear algebraic tasks in learning theory.
For example, $\lra$ is an effective tool in reducing the memory burdens of kernel methods. In kernel methods,  we map the data into a kernel-induced Hilbert space.
Given $n$ samples, this requires the calculation of an $n \times n$ symmetric, positive semi-definite kernel matrix, which requires quadratic space and mostly cubic time. This can be very demanding for large problems, impeding its practical deployment. This is where $\lra$ is helpful. Williams and Seeger~\cite{WS00} showed that the spectra of kernel matrices decay rapidly. This led to the following approach: compute the $\lra$ of the kernel matrices and perform computation on those matrices. This simultaneously brought improvement in many other models, like support vector machine, kernel Fisher discriminant analysis and kernel principal component analysis.
Similarly, \linear, which is used to predict the output for a new data based on the previous data, has played pivotal role in machine learning. Examples include, {\em capital asset pricing model} which is used to predict demands, supplies, and investment, 
{\em Reinforcement Learning} to approximate value functions, and learning parameters of a model with linear dynamics (see, for eg., Walsh  {\it et al.}~\cite{WSDL09}). Likewise, \mult \ has found many roles in machine learning other than modelling multivariate regression as a transposed \mult \ problem. For example, Mossel {\it et al.}~\cite{MOS03} showed how to improve run time of learning juntas using matrix multiplication. 

\paragraph{Motivation of this work.} The data in the examples mentioned above and on which we perform the learning  contain sensitive information and takes a lot of space. As a result, privacy and memory constraints are as important as correct computation. The privacy issue has been recently exemplified by the deanonymization of Netflix datasets, which was cited as one of the reasons to study differentially private low-rank approximation by~\cite{HR12,HR13}. Similarly, data used in genetics engineering and  finance have natural privacy concerns. This raises the question of whether one can perform all these tasks while giving a robust guarantee of  privacy, like {\em differential privacy}. When privacy is not a concern, there are many algorithms which use almost optimal space and one-pass over the input matrix (see, for example, Kane and Nelson~\cite{KN14}).  
On the other hand, all the known private algorithms that achieve small additive error use multiple passes over the matrix.  For example, the traditional {\em Krylov subspace iteration} method (on which some of the recent works like~\cite{Hardt13,HR13,KT13} are based) requires a lot of space and multiple passes over the input matrix. {\em The main focus of this work is to perform linear algebraic computation while preserving privacy of the data with the additional challenge that the data is received online and we can use sub-linear space.}  

\paragraph{First sketch based differentially private mechanisms.} 
A natural approach to perform linear algebraic tasks on online data while preserving privacy is to generate and store a sub-linear space data structure that can be used to perform these tasks without leaking privacy.  
In the non-private setting,  the standard techniques  either perform {\em random sampling}  or generate a {\em random sketch} of the input data. Dwork  {\it et al.}~\cite{DNPRY} showed that known (non-private) random sampling based algorithms for certain statistical queries can be made differentially private. They also gave an impossibility result for private analogues of  {\em sketch based approaches} for specific ``statistical queries."
This raises doubts over the applicability of sketch based approach in privacy. In this paper, {\em we show the first set of positive results for sketch based approach on streamed data. We give almost optimal space data-structures that can compute $\lra$, \linear, and \mult \ while preserving differential privacy.} Notably, we match the best known space bounds for all these tasks in the non-private setting. We remark that our results does not contradict the impossibility result of Dwork  {\it et al.}~\cite{DNPRY}, which holds for cleverly constructed statistical queries. 



\paragraph{\scshape Privacy model used in this paper.} There are two commonly used notions of differential privacy: {\em event level privacy}, where guarantees are at the granularity of individual records in the datasets, and {\em user level privacy}, where guarantees are at the granularity of each user whose data is present in the dataset. Dwork {\it et al.}~\cite{DNPR10} showed that it is impossible to obtain any non-trivial result with respect to the user level privacy on a streamed data. Therefore, in this paper, we use event level privacy. More specifically, we say two matrices $\bA_1$ and $\bA_2$  with same rows and columns are {\em neighbouring} if the matrix $\bA_1- \bA_2$ is a rank-$1$ matrix  and $\| \bA_1 - \bA_2 \| \leq 1$. This notion was also used recently by \cite{HR12,HR13,Hardt13} for differentially private low-rank approximation and by Blocki {\it et al.}~\cite{BBDS12,BBDS13} and Upadhyay~\cite{Upadhyay13} for certain statistical queries. We also restrict how the mechanism receives the data in the sense that the data matrix is streamed online and arrives either row-wise or column-wise.  We call two data streams neighbouring if they are formed by streaming entries of neighbouring matrices. With this notion of neighbouring datasets, we use the following definition for privacy.

\begin{definition} \label{def:approxdp}
	A randomized algorithm $\mathfrak{M}$ gives {\em $(\varepsilon, \delta)$-differential privacy}, if for all neighbouring 
data-streams $\bA_1,\bA_2$ and all $S \subset Range(\mathfrak{M})$, 
		$ \p[\mathfrak{M}(\bA_1) \in S] \leq \exp (\varepsilon) \p[\mathfrak{M}(\bA_2) \in S] + \delta, $ where the probability is over the coin tosses of $\mathfrak{M}$. 
\end{definition}

The main focus of this study is the space requirement of the privacy-preserving data-structures.

\subsection{\scshape Problem Statements and our Results} \label{sec:prob}
In this section, we give the formal description of the problems we investigate in this paper. 
The performance of a streaming algorithm is measured by three basic factors: the number of passes over the data stream, the space used by the data-structure, and the time taken to update the data-structure. All our  private mechanisms for performing linear algebraic tasks are single-pass and achieve almost optimal space bound  for one-pass algorithms. Our private sketch are linear; therefore, {\em our mechanisms extends naturally to turnstile  updates}. We reserve the letter $n$ for the number of rows and $d$ for the number of columns of a private matrix. We assume $d < n$.
 For {\em bit complexity}, we use the convention of Clarkson and Woodruff~\cite{CW09}, i.e., the entries of a matrix can be represented by $\kappa=\log(nd)$ bits. 

\medskip
\noindent {\scshape Low-rank Approximation.} 
We start with the problem statement of $\lra$.   
\begin{prob} {\em ($(\alpha, \beta, \tau)$-$\lra$).} \label{prob:low}
Given an $n \times d$ matrix $\bA$ and a target rank $k$, construct a matrix $\bPsi$ with $k$ 
orthonormal columns such that $\| \bA - \bPsi \bPsi^{\mathsf T}  \bA \|_N \leq (1+\alpha) \min_{{\sf rank}(\bA_k) \leq k} \| \bA- \bA_k \|_N + \tau$ with probability at least $1-\beta$ where $N$ represents either Frobenius or spectral norm. 
\end{prob}

The best known space lower bound in the non-private setting for $(\alpha, 5/6, 0)$-$\lra$ is $\bOmega(nk\alpha^{-1}\kappa)$ due to Clarkson and Woodruff~\cite{CW09}.
Kane and Nelson~\cite[Th 6.2]{KN14} improved the earlier analysis of Clarkson and Woodruff~\cite{CW09} to show a one-pass non-private streaming algorithm for approximate $\lra$ with respect to the Frobenius norm with row/column-wise updates where one maintains a data-structure in the form of a sketch of size $O(k\alpha^{-1}(n + d) \kappa \log(1/\beta) )$.

The first approach to prove differential privacy using the sketch generation algorithm of Clarkson and Woodruff~\cite{CW09} is to use additive noise mechanisms. There are two main drawback of this approach. A straightforward analysis along the line of~\cite{HR12} with no coherence assumption leads to an additive error, $\tau$, of order $k^{3/2}$ for the Frobenius norm approximation. The second drawback is that it only leads to approximation when the approximation norm is Frobenius. In this paper, we use random Gaussian matrices to generate the sketch and prove the following in~\secref{lsi} (see~\thmref{lsi} for the precise statement). 
\begin{quote} {\bf Theorem (Informal).}
There is an $O(k\varepsilon^{-1}(n + d) \kappa)$ bits data structure that could be used to publish $(\varepsilon, \delta)$-differentially private $k$-rank approximation of an $n\times d$ private matrix in a single pass  with $ \tau \leq O ( \sqrt{nk \ln (2/\delta)}/\varepsilon)$ for the Frobenius and $ \tau \leq O ( \sqrt{nk \ln (2/\delta)}/\varepsilon)$ for the spectral norm. 
\end{quote}

Note that our data-structure provides $\lra$ in both the spectral and Frobenius norm with the same space bound as in the non-private setting. The reason behind this is simple. The bound on the single pass algorithm of \cite{CW09,KN14} and the two-pass algorithm of Sarlos~\cite{Sarlos06} use their bound for \mult \ (see Problem~\ref{mult}), and, therefore, only achieves a bound on the Frobenius norm. On the other hand, we use perturbation theory which gives us a unified bound for the both the norms. Notably, our proof does not rely on our bound in \mult \ (see our result for Problem~\ref{mult}). 

\begin{table} [t]
{
\small{
 \begin{center}
\begin{tabular}{|c|c|c|c|c|c|c|}
\hline
Method & Additive noise ($\tau$)  &  Privacy Notion & $\#$ Passes   \\ \hline
Chaudhary {\it et al.}~\cite{CSS12} & $O(nk/\varepsilon)$ & Spectral norm & $k$ \\ \hline 
Hardt and Roth~\cite{HR12}			& $\frac{\sqrt{kn}\log(k/\delta)}{\varepsilon} + \sqrt{\frac{\mu \|A\|_F\log(k/\delta)}{\varepsilon}}$ 	& Event level & 2   \\ \hline
Hardt and Roth~\cite{HR13}	& $O(\frac{k^2}{ \varepsilon} \sqrt{ (\mathsf{rk}(A)\mu + k\log n) \log \paren{ \frac{1}{\delta}}} \log n)$ 	& Event level  & $k \sqrt{\log \sigma}$  \\ \hline
Kapralov and Talwar~\cite{KT13} 		& $ O(dk^3/(\varepsilon \gamma^2))$ 									& Spectral norm & $k$ \\ \hline
Hardt and Price~\cite{Hardt13}				& $\frac{\sigma_1 \sqrt{kn \mu \log(1/\delta) \log \paren{{n}/{\gamma}} \log \log  \paren{{n}/{\gamma}}}}{\varepsilon \gamma^{1.5} \sigma_k}$ & Event level  & $k \sqrt{\log \sigma}$ \\ \hline
Dwork {\it et al.}~\cite{DTTZ14}	& $O((k \sqrt{n} \ln(1/\delta))/\varepsilon) + \tilde{O}(\sqrt{k^3n^{3/2}}/\varepsilon^2)$ 	 & User level &$1$ \\ \hline
This paper (\thmref{lsi}(i))					& $O ({ \sqrt{nk \ln (k/\delta)}}/{\varepsilon} )$  &Event level  & $1$  \\ \hline
This paper (\thmref{lsi}(ii))					& $O ({ \sqrt{nk \ln (k/\delta)}}/{\varepsilon} )$  &Event level & $1$ \\ \hline
\end{tabular}
\caption{Comparison Between our Mechanism and Previous Mechanisms for Differentially Private $k$-Rank Approximation of an $n \times d$ rank-$\mathsf{rk}(A)$ matrix A.  $\mu$ denotes the coherence of $\bA$, $\gamma=\paren{\sigma_k/\sigma_{k+1}}-1$.~\cite{HR12} and~\thmref{lsi}(i) bounds the Frobenius norm while the other entries bounds the Spectral norm.} \label{table}
 \end{center}
}}
\end{table}

\noindent {\sc Comparison and tightness.} Let $\sigma_1, \cdots, \sigma_{\mathsf{rank}(\bA)}$ be the singular values of a matrix $\bA$. We compare our result for $\lra$ with the previous works in Table~\ref{table}. Due to lack of space, we defer the detail comparison to Appendix~\ref{sec:comparison}. Here we just state few main points. Works like~\cite{CSS12,DTTZ14,KT13} consider different notion of neighbouring datasets, so our results are incomparable to theirs.
Apart from the result in this paper, only Dwork {\it et al.}~\cite{DTTZ14} give $\lra$ on streamed data though under extra assumptions like {\em normalized row} assumption and a lower bound on the optimal value. Also, we do not make any {\em low-coherence} assumption. Therefore, if we set $\mu=n$ (for worst-case) to compare our results, then Table~\ref{table} shows that we achieve best bound on $\tau$ with the same privacy model. In fact, we achieve almost optimal $\tau$ for spectral norm (see,~\cite[Thm. 6.1]{HR13}). Our result for Frobenius norm is also tight due to the result of Blum {\it et al.}~\cite{BDMN05}. 

\medskip
\noindent {\scshape Linear Regression.} We start by giving the problem statement. 

\begin{prob} {\em ($(\alpha, \beta, \tau)$-\linear).}
Given an $n\times d$ matrix $\bA$ and a $m$ set of $n\times 1$ column 
vectors $\mathbf{B}=\{ \mathbf{b}_1, \cdots, \mathbf{b}_{m}\}$, output a set of vectors $\mathbf{X}=\{\bx_1, \cdots \bx_m\}$ all in $\R^d$
so that for all $i \in [m]$, $\| \bA\bx_i - \mathbf{b}_i \|_F \leq (1+\alpha) \min_{\by \in \R^{d \times 1}} \|\bA\by-\mathbf{b}_i \|_F + \tau$ with probability at least $1-\beta/ m$.
\end{prob}

The best known space lower bound for $(\alpha, 5/6, 0)$-\linear \ in the non-private setting is $\Omega(d^2\alpha^{-1} \kappa )$ due to~{\cite{CW09}}. Kane and Nelson~\cite[Th 6.2]{KN14} improved the analysis of Clarkson and Woodruff~\cite{CW09} to  gave a $(d^2\alpha^{-1} \kappa \log(1/\beta) )$ bits data-structure that can answer $(\alpha, \beta, 0)$-\linear \ in the non-private setting. We show that their space bound can be achieved in the private setting at the expense of a small additive error. Informally, we show the following in~\secref{linear} (see~\thmref{linear} for the precise statement).

\begin{quote} {\bf Theorem (Informal).}
There is a data-structure that uses $O\paren{ d^2 \alpha^{-1}\kappa \log\paren{{1/\beta}}}$ bits of space to compute $(\alpha,\beta,O(s^2\alpha \sqrt{n}))$-\linear \ while  providing $(\varepsilon,\delta)$-differential privacy, where  $s =  \sqrt{ \frac{16d \log\paren{{1}/{\beta}}  \ln \paren{{2}/{\delta}}}{\alpha \varepsilon^2}  } \ln \paren{ \frac{16d \log(1/\beta)}{\alpha \delta} }$. 
\end{quote}
\linear \ has been also studied in the {\em local privacy model} by Duchi {\it et al.}~\cite{DJW13} and in the {\em online private learning model} by Jain {\it et al.}~\cite{JKT12} and Thakurata and Smith~\cite{TS13}. These models are different from ours; therefore, our results are incomparable to theirs. 

\medskip
\noindent {\scshape Matrix Multiplication.} We start by giving the problem statement.  
\begin{prob} \label{mult}
	{\em ($(\alpha, \beta, \tau)$-\mult).} An $n \times d_1$ matrix $\bA$ and $n \times d_2$ matrix $\mathbf{B}$ are given.  Output a matrix C so that $\| \bA^{\mathsf T} \mathbf{B} - \mathbf{C} \|_F \leq \alpha \|\bA \|_F \cdot \| \mathbf{B} \|_F + \tau$ with probability at least $1-\beta$.
\end{prob}

Let $d=\max \{d_1,d_2\}$. The best known space lower bound for $(\alpha, 5/6, 0)$-\mult \ in the non-private setting is $\Omega(d \alpha^{-2} \kappa )$ due to Clarkson and Woodruff~\cite{CW09}. Kane and Nelson~\cite[Th 6.2]{KN14} improved the analysis of Clarkson and Woodruff~\cite{CW09} to give a $O\paren{d \alpha^{-2}\kappa \log\paren{{1/\beta}}   }$-bits data-structure that can solve $(\alpha, \beta, 0)$-\mult \ in the non-private setting. In~\secref{prod}, we show that this bound can be achieved in the private setting with a small additive error (see~\thmref{prod} for the precise statement).

\begin{quote} {\bf Theorem (Informal).}
Given conforming matrices $\bA$ and $\mathbf{B}$. Let $r=O( \log(1/\beta)/\alpha^2)$ and $s =  {\sqrt{16r  \ln(2/\delta)}} \varepsilon^{-1} \ln(16r /\delta)$. For  large enough $d=\max \{d_1,d_2\}$, there is data structure which maintains a sketch of size $O\paren{d \alpha^{-2}\kappa \log\paren{{1/\beta}}   }$, computes $(\alpha, \beta, O(s^2\alpha \sqrt{n}))$-\mult \ and provides $(\varepsilon,\delta)$-differential privacy. 
\end{quote}

\Jcom
{Let $\Phi'=\Omega \Omega^{\mathsf T} A$. The right calculation gives the PDF of the published matrix as 
$$p_{\Phi'}(\Phi')  \propto p_\Phi (\Phi' A^{-1} )  \propto \exp( \mathsf{Tr} (-\Sigma^{-1} Y A^{-1}/2) = \exp( \mathsf{Tr} (-A^{-1} \Sigma^{-1} Y)/2).$$ 
This results in  $\mathsf{Tr} ((A^{-1} E \widetilde{A}^{-1})\Phi)$ at the end of the page and eq 7 would change to 
 $$\mathsf{Tr} (\bra{\mathbf{g}_j} V \Sigma^{-1} U^{\mathsf{T}}  \ket{\mathbf{v}} \bra{ \mathbf{e}_i} \widetilde{U} \Lambda^{-1} \widetilde{V}^{\mathsf{T}}\ket{\mathbf{g}_j}).$$
We can simply evaluate the two terms $\bra{\mathbf{g}_j} V \Sigma^{-1} U^{\mathsf{T}}  \ket{\mathbf{v}}$ and $\bra{ \mathbf{e}_i} \widetilde{U} \Lambda^{-1} \widetilde{V}^{\mathsf{T}}\ket{\mathbf{g}_j}$ as in the present text to get 
$\mathsf{Pr} [(7) \leq 2(1/\sigma_{\mathsf{min}} + 1/\sigma_{\mathsf{min}}^2 \ln (4/\delta_0)] \geq 1-\delta$. 
The first part of (6) is straightforward as we get a bound of $\exp(\pm \varepsilon/2r)$ instead of $\exp(\pm \varepsilon/r)$. 
Both of these are still under the permissible range of $\sigma_{\min}$ for $PSG_2$. Therefore, the privacy guarantee still holds.  Unless we are missing something, A-(-A) would be rank n matrix; however, our definition of neighbouring matrices also requires that the differing row or column has norm 1! The privacy leak would depend on the edit distance between A and -A under this extra condition.
}

\noindent {\sc Space optimality of the data structures.} In Table~\ref{table:compare}, we give the best known lower bound for the space required for all the problems we study in this paper, that by the best known data-structures in the non-private setting, and by our private data-structures. The parameters used in Table~\ref{table:compare}  are the same as in Table~\ref{table}. As one can see, we achieve almost optimal space data-structures for all the three problems. Note that, we explicitly require $\bOmega$ to be stored for the mechanism of $\lra$; in the other two cases, $\bOmega$ can be picked from the distribution on the fly.  We note that stronger lower bounds on space are achievable for each of the problems considered because of the non-zero additive error; however, the bounds stated in the table are presently the best known lower bounds that we can use to make any sort of optimality comparison. 
\begin{table} [t]
\begin{center}
{
\small{
\begin{tabular}{|c|c|c|c|}
\hline
Problem & Lower Bound & Non-private Algorithm  &  This paper \\ \hline
$(\alpha,\beta,\tau)$-\mult \ & $\Omega \paren{ d\alpha^{-2} \kappa}$ & $\tilde{O}(d\alpha^{-2} \kappa \log (1/\beta) )$ & $\tilde{O}(d\alpha^{-2} \kappa \log (1/\beta) )$   \\ \hline
$(\alpha,\beta,\tau)$-\linear \ & $\Omega \paren{{ d^2 \alpha^{-1} \kappa} }$& $\tilde{O}\paren{ d^2\alpha^{-1}\kappa  \log (1/\beta) }$  & $\tilde{O}\paren{ d^2\alpha^{-1}\kappa  \log (1/\beta) }$   \\ \hline
Frobenius norm $(\alpha,\beta,\tau)$-$\lra$ & $\Omega(nk\kappa / \alpha)$ & $\tilde{O}(k\alpha^{-1}(n + d) \kappa )$ & $O(k\varepsilon^{-1}(n + d) \kappa)$  \\ \hline
Spectral norm $(\alpha,\beta,\tau)$-$\lra$ & $-$ & $-$ & $O(k\varepsilon^{-1}(n + d)\kappa )$  \\ \hline
\end{tabular}
\caption{Our Results with Respect to the Best Known Space Bounds in Non-private Setting. } \label{table:compare}
}}
\end{center}
\end{table}

One can also implement our mechanisms  {\em as distributed algorithms}, a desirable feature as argued by~\cite{BNO08}. This is because all the operations used in our mechanisms have efficient distributed algorithms (Jacobi method for singular value decomposition, Cannon~\cite{Cannon69}'s algorithm for multiplication, and GMRES~\cite{SS86}'s residual method).
\subsection{\scshape Our Techniques}
We start by giving an overview of the generic private sketch algorithm and then discuss the ideas used in this work  which are different from other related works. 

\medskip
\noindent {\scshape Private-sketch Generation.} In order to get a better bound on the additive noise, we devise a private-sketch generation ($\mathsf{PSG}$) mechanism to generate a private sketch of the input stream. We reduce the privacy guarantee to maintaining a spectral property of the streamed matrix. At a high level, all our mechanisms use this basic mechanism while maintaining the spectral properties of the private matrix. We note that the affine transformation of Blocki {\it et al.}~\cite{BBDS12, BBDS13} and Upadhyay~\cite{Upadhyay13} could be also used to maintain the required spectral properties, but this leads to a large additive error. Moreover, as noted by Blocki {\it et al.}~\cite{BBDS12}, it does not guarantee $\lra$. For these reasons, we use a different method. 

\medskip
\noindent {\scshape Mechanism and analysis for $\lra$.} We follow two-steps to compute a $\lra$: (i) compute a projection matrix ({\em range finding} step), (ii) compute a $\lra$ by operating the projection matrix on the input matrix ({\em projection} step). In the most naive form, both the steps require the input matrix, and hence two-passes are required. The first observation is that the mechanism for $\mathsf{PSG}$ already gives  considerable improvement in the range finding step. {This by itself does not give $\lra$ as the singular values of the projection matrix and private matrix are not comparable at this stage.} We need to use the projection step. The second observation is that, by a clever use of linear algebra,  information gathered in the range-finding step using a random Gaussian matrix can be used to emulate the projection step without using the input matrix explicitly. However, we need to reuse the random Gaussian matrix. This complicates the privacy proof (discussed below in more detail). We also use the idea of Hardt and Price~\cite{Hardt13} to use an oversampling parameter $p$ to get a sharper bounds for both the Frobenius and spectral norm.  

\medskip
\noindent {\scshape Differences in techniques to compute $\lra$.} Our mechanism for $\lra$ is markedly different from the recent private mechanisms~\cite{CSS12,DTTZ14, Hardt13, HR12, HR13,KT13}. Among all these mechanisms, only Dwork {\it et al.}~\cite{DTTZ14} computes $\lra$ privately in an online manner by using binary tree technique of  Dwork {\it et al.}~\cite{DNPR10}.  We differ at a basic level from these mechanisms: all these works perturb the output by adding noise to it, while {\em we perturb the input matrix and then multiply noise matrix}.  
Our mechanism uses random Gaussian matrices (privacy proof holds only for Gaussian matrix); therefore, we use different tricks, like reusing Gaussian matrix, as mentioned in the last paragraph.

\medskip
\noindent {\scshape Differences from previous analyses of $\lra$.} Our analysis has some interesting features. As mentioned above, we reuse the Gaussian matrix to get a mechanism that uses only one-pass over the input matrix. In general, reusing randomness can result in privacy breach. Fortunately, the Gaussian matrix is reused in a specific manner for which we prove that privacy holds under certain spectral property of the input matrix. We believe this can be of independent interest. Our mechanism works by maintaining the required singular value for the input matrix. This causes the additive error. The analysis used in this paper to bound the error  differs a lot from  the analyses of Clarkson and Woodruff~\cite{CW09} and Sarlos~\cite{Sarlos06}.  They use the trick that a good bound on \mult \ allows a bound on $\lra$. This limits its applicability to the case when the approximation metric is the Frobenius norm. Moreover, Sarlos~\cite{Sarlos06} use two-passes over the input matrix. We use perturbation theory, which  allows us to give bounds for both the norms in an unified manner. We believe this can be of independent interest.  

\medskip
\noindent {\scshape Other results: Mechanism for \mult \ and \linear.} At a high level, our mechanisms for $(\alpha, \beta, \tau)$-\mult \  and $(\alpha, \beta, \tau)$-\linear \  are private analogues of Clarkson and Woodruff~\cite{CW09}. Their algorithm use  {\em tug-of-war matrices} with Rademacher entries (this helps in improving the update time). Adding Gaussian noise to ensure differential privacy amounts to a large additive error. We use {\sf PSG} mechanism.  For the privacy proof to go through, we need to lift the singular values of the input matrix. Our choice to lift the singular values is constrained by keeping a check on $\tau$ as well as to keep the mechanism one-pass.  
In order to use the proofs of Kane and Nelson~\cite{KN14} to give an approximation bound, we use an analogous variance bound on random Gaussian matrix (since privacy does not necessarily hold for random matrices used by Kane and Nelson~\cite{KN14}). 

\paragraph{\scshape Related Works.}
The first formal definition and mechanism for differential privacy was  given by Dwork {\it et al.}~\cite{DMNS06}. 
Since then, many mechanisms for preserving differential privacy have been proposed in the literature (see, Dwork and Roth~\cite{DR13}). 
The literature on non-private streaming algorithms is so  extensive that we cannot hope to cover it in any detail here. In the private setting, Dwork  {\it et al.}~\cite{DNPRY} studied {\em pan-privacy}, where the internal state is known to the adversary. 
Subsequently, there have been some works on online private learning, like Dwork  {\it et al.}~\cite{DTTZ14}, Jain {\it et al.}~\cite{JKT12}, and Thakurata and Smith~\cite{TS13}, for various tasks.  There are some recent works on differentially private $\lra$ as well. Blum  {\it et al.}~\cite{BDMN05} first studied this problem and gave a simple ``input perturbation" algorithm that adds noise to the covariance matrix. This was improved by Hardt and Roth~\cite{HR12} who studied $\lra$ for the Frobenius norm under the low coherence assumption. Kapralov and Talwar~\cite{KT13} and Chaudhary  {\it et al.}~\cite{CSS12} studied the spectral $\lra$ of a matrix by giving a matching upper and lower bounds for privately computing the top $k$ eigenvectors of a matrix. Hardt and Roth~\cite{HR13} and Hardt and Price~\cite{Hardt13} improved their noise bound by proposing robust private subspace iteration mechanism. Recently, Dwork  {\it et al.}~\cite{DTTZ14} revisited randomized mechanism to give a tighter bound and used it to give an private online learning algorithm under a normalized row assumption.

\section{ Notations and Basic Preliminaries}\label{sec:prelims}
We reserve the letters $\bA$ and $\mathbf{B}$ for  private input matrices, and $\bOmega$ for a random  Gaussian matrix. For an $n \times d$ matrix $\bA$, we let  $\bA_{i:}$ denote the $i$-th row of  $\bA$, $\bA_{:j}$  denote the $j$-th column of $\bA$, and $\bA'$  denote the symmetric matrix $\begin{pmatrix} 0 & \bA \\ \bA^{\mathsf T} & 0 \end{pmatrix}$ corresponding to $\bA$. We let $\bA_t$ denote the matrix received after $t$ time epochs. 
The singular value decomposition ($\mathsf{SVD}$) of $\bA$ is $\bA=\bV \bSigma \bU^{\mathsf T}  $, where $\bU$ and $\bV$  are left and right eigenvectors of $\bA$,  and $\bSigma $ is a diagonal matrix. The entries of $\Lambda $ are called the {\em singular values} of $\bA$. Since $\bU$ and $\bV$ are unitary 
matrices, one can write $\bA^i = \bV \bSigma^i \bU^{\mathsf T}  $ for any real value $i$. We let  $\mathsf{rank}(\bA)$ denotes the rank of the matrix $\bA$ and $\sigma_i(\bA)$ its singular values. Where it is clear from context, we  write $\sigma_i$ for the singular values.

We use various matrix norm. We use the notation $\| \cdot \|_F$ for Frobenius norm. A Frobenius norm for a matrix $\bA=(a_{ij})_{i\in [n],j\in[d]}$ is defined as following $\|\bA\|_F = \sum_{ij} |a_{ij}|^2$. For a matrix $\bA$, we let $\|\bA\|_2$ denote the $2$-norm, i.e., $\max_{\bx \in \R^d} \| \bA \bx\|_2/\|\bx\|_2$.
s
When we wish to refer to both the Frobenius and the spectral norm, we overload the symbol $\| \cdot \|$ and drop the subscript.  We let $\mathbf{e}_1, \cdots , \mathbf{e}_d$ denote the standard basis vectors in $\R^d$. We use  bold face symbols to denote vectors and ${\mathbf{0}^n}$ to denote an $n$-dimensional $0$-vector. For a matrix $\mathbf{M}$, we write $\mathbf{M} \succeq 0$ if all its eigenvalues are non-zero. 
In the course of this paper, we use many standard results from linear algebra and random matrices.  
\begin{extended}
Due to lack of space, we defer the detail discussion in~\appref{facts}.
\end{extended}

\begin{full}
\paragraph{\scshape Linear Algebra.}
Our analysis make extensive use of linear algebra and statistical properties of Gaussian distribution. We give an exposition to the level required to understand this paper.

\begin{lemma} \label{lem:conform}
	If matrix $\bA$ and $\mathbf{B}$ are conforming, then $\|\bA \mathbf{B}\|_F \leq \|\bA\|_2 \|\mathbf{B} \|_F.$
\end{lemma}

\begin{lemma} \label{lem:hermitian} 
	Let $\bA$ and $\mathbf{B}$ be Hermittian matrices. If $\bA$ and $\mathbf{B}$ differ in at most one row of Euclidean norm $1$, then $\tr(\bA^{\mathsf T}\bA) - \tr(\mathbf{B}^{\mathsf T} \mathbf{B}) \leq 2$.
\end{lemma}

\paragraph{\scshape Univariate and Multivariate Gaussian Distribution.}
A random variable, $X$, distributed according to a Gaussian distribution has the probability density function, $ \PDF_X(x) = \frac{1}{\sqrt{2\pi \sigma}} \exp \paren{- \frac{(x-\mu)^2}{2\sigma^2}}. $ We denote it by  $X \sim 
\cN(\mu, \sigma^2)$. The Gaussian distribution is invariant under affine transformation, i.e., if $X 
\sim \cN(\mu_x, \sigma_x)$ and $Y \sim \cN(\mu_y, \sigma_y)$, then $Z=aX +bY$ has the distribution $Z \sim \cN(a \mu_x + b \mu_y, a
\sigma_x^2 + b\sigma_y^2)$.

The multivariate Gaussian distribution is a generalization of univariate Gaussian distribution. Given a $m$ dimensional multivariate random 
variable, $X \sim \cN(\mu, \bSigma)$, the $\PDF$ of a multivariate Gaussian is given by $ \PDF_\bX(\bx) := \frac{1}{\sqrt{(2 \pi)^{\mathsf{rank}(\bSigma)} \mathsf{Det}(\bSigma)}} \exp \paren{-\frac{1}{2} \bx^{\mathsf T}    \bSigma^\dagger   \bx}$  with mean $\mu \in \R^m$ and covariance matrix $\bSigma =\E[(X- \mu)(X-\mu)^{\mathsf T}  ]$.   If $\bSigma$ has a non-trivial kernel space, then the $\PDF$ is undefined. However, in this paper, we only need to compare 
the probability distribution of two random variables which are defined over the same subspace. Therefore, wherever required, we restrict our attention to the 
(sub)space orthogonal to the kernel space of $\bSigma$. 
Multivariate Gaussian distribution maintains many key properties of univariate Gaussian distribution. For example, any (non-empty) subset of 
multivariate Gaussians is a multivariate Gaussian and  linear functions of multivariate Gaussian random variables are multivariate Gaussian random variables, i.e., if 
$\by = \bA \bx+\mathbf{b}$, where $\bA \in \R^{n \times n}$ is a non-singular matrix and $\mathbf{b} \in \R^n$, then $\by \sim \cN 
(\bA \mu+\mathbf{b},\bA \bSigma \bA^{\mathsf T}   )$. 

We use the following properties of Gaussian matrices, a proof of the first lemma could be found in~\cite{Tao}, while for the rest two, one can refer to~\cite{Muirhead}.
\begin{lemma} \label{lemma:random1} 
	Let $\bOmega$ be a  Gaussian matrix. Then  
	$ \E[\| \bA \bOmega \mathbf{B} \|_2] \leq \| \bA \|_2 \| \mathbf{B} \|_F + \|\bA \|_F \| \mathbf{B} \|_2 $ for fixed matrices $\bA$ and $\mathbf{B}$.
\end{lemma}

\begin{lemma} \label{lemma:random2} 
	Let $\bOmega$ be a $n \times (k+p)$ Gaussian matrix. Then 
	\begin{align*} \E[\| \bOmega^{-1} \|^2_F ] = \sqrt{\frac{k}{p-1}} \qquad \text{and} \qquad \E[\|  \bOmega^{-1} \|_2] \leq \frac{e \sqrt{k+p}}{p}. \end{align*}
\end{lemma}

\begin{lemma} \label{lemma:wishart}
	Let $\bOmega$  be a random $n \times (k+p)$ Gaussian matrix whose entries are picked from the distribution $\cN(0,1)$. Then $\E [\tr((\bOmega^{\mathsf T} \bOmega)^{-1}) ]= k/(p-1)$.
\end{lemma}

\begin{theorem} \label{thm:JL} {\em (Johnson-Lindenstrauss lemma)}
	Fix any $\eta <1/2$ and let $m$ be a positive integer. Let $\bOmega$ be a $k \times n$ matrix with entries picked from a Gaussian distribution $\cN(0,1)$, where $k \geq 4(\alpha^2/2-\alpha^3/3)^{-1} \ln m$. Then for any $m$ unit vector set  $S$ in $ \R^n$
	\begin{align*}\forall \bx \in S, \p_\mathbf{M} \sparen{ \| \bOmega \bx \|_2 \in (1 \pm \alpha) \|\bx\|_2 } \geq 1 - 2 \exp (-\alpha^2 k/8). \end{align*}
\end{theorem}

\paragraph{\scshape Differential Privacy.}
We use the following in our analysis explicitly or implicitly.
\begin{theorem} {\em (\cite{DRV10}).} \label{thm:DRV}
	Let $\varepsilon, \delta \in (0,1)$, and $\delta'>0$. If $\cK_1, \cdots , \cK_\ell$ are each $(\varepsilon, \delta)$-differential private mechanism, then the mechanism $\cK(D):= (\cK_1(D), \cdots , \cK_\ell(D))$ releasing the concatenation of each algorithm is $(\varepsilon', \ell \delta+\delta')$-differentially private for $\varepsilon' < \sqrt{2\ell \ln (1/\delta')}\varepsilon + 2\ell \varepsilon^2$.
\end{theorem}
\begin{lemma} \label{lem:post}
Let $M(D)$ be a $(\varepsilon, \delta)$-differential private mechanism for a database $D$ , and let $h$ be any function, then any mechanism $M':=h(M(D))$ is also $(\varepsilon,\delta)$-differentially private for the same set of queries.
\end{lemma}

\end{full}

\section{Differentially Private Sketch Generation} \label{sec:main}
{
\begin{figure} [t]
\fbox{
\begin{minipage}[l]{6.3in}
\small{ On input a streamed column $\mathbf{v} \in \R^{2 \widetilde{n}}$ of the  private matrix, parameters $r,\tilde{n}$, the  mechanism samples an $r \times 2\tilde{n}$ random matrix $\bOmega$, and
\begin{description} 
\item [${\sf PSG}_1$:] Compute $\bY_{\mathbf{v}}=\bOmega \mathbf{v}$, and return $\bY_{\mathbf{v}}$.
\item [${\sf PSG}_2$:] Compute $\bY_{\mathbf{v}}=\bOmega^{\mathsf T} \bOmega\mathbf{v}$, and return $\bY_{\mathbf{v}}$.
\end{description}	 }
\end{minipage}
}
\caption{Private Sketch Generation ($\mathsf{PSG}$) Algorithm} \label{fig:psg} 
\end{figure}
}
We study differential privacy in the well known streaming model of computation~\cite{AMS99}. We present it at the level required to understand this paper (a more formal definition can be found in~\cite{AMS99}). This model has three entities: a stream generator $\cS$, a (database) curator $\cK$, and a query maker $\cQ$.
	 $\cS$ starts the process at time $t=0$ and the curator initializes its data structure to $\cD_0$. Thereafter, the curator is allowed only one-pass over the input matrix, i.e., it can access any entry of the data-base during exactly one  time epoch, and 
	 update its data structure to $\cD_t$  using $\cD_{t-1}$ and the newly accessed data-points of the matrix. 
	 At any time, $t$, the query maker $\cQ$ makes a query $q$. The curator responds with  $q(\cD_{t})$.
The streaming model  has a resource bound on the curator. A curator is only allowed to use  total time {\em polynomial in the size of the data-base} to construct the data structure, where the size of data-structure should be  {\em sub-linear in the size of the data-base}. 
 For {\em differential privacy, we further require that the response of $\cK$  to the query of $\cQ$ should satisfy Definition~\ref{def:approxdp} with respect to two neighbouring streams.} Differential privacy on streaming data has been also studied as private online learning by~\cite{JKT12} and~\cite{TS13}, where the emphasis is on learning about the streamed input -- one bears a regret on a hypothesis evaluated against a data point which is not yet streamed. Due to lack of space, we do not delve into the detail comparison of the two models. Our main privacy result is as follows.
\begin{theorem} \label{thm:PSG}
(i) If the singular values of the  matrix whose columns are streamed to $\mathsf{PSG}_1$ are at least $\sigma_\mathsf{min}:={ \paren{ 4\sqrt{r \log (2/\delta)} \log(r/ \delta)}/{\varepsilon}}$, then $\mathsf{PSG}_1$ preserves $(\varepsilon, \delta)$-differential privacy. (ii) If the singular values of the  matrix whose columns are streamed to  $\mathsf{PSG}_2$ algorithm are at least  $\sigma_\mathsf{min}:={ \paren{4r \log(r/ \delta)}/{\varepsilon}}$, then $\mathsf{PSG}_2$ preserves $(\varepsilon, \delta)$-differential privacy. 
\end{theorem}
The proof of part (i) needs some care due to subtleties mentioned later in this section, but the overall outline follows the idea of Blocki {\it et al.}~\cite{BBDS12}. However, the proof of part (ii) is very different and more involved.  We show that for a streamed matrix $\bA$, the $\PDF$ of the published matrix using $\mathsf{PSG}_2$ is 
\begin{align} \frac{ \exp (-\tr (\bA^{-1} \bPhi )/2) \Delta(\bPhi)^{(n-r-1)/2}} {2^{rn/2} \pi^{r(r-1)/4} \Delta(\bA)^{r/2} \prod_{i=1}^r \Gamma((n-i+1)/2)}, \label{eq:pdf} \end{align} 
where $\Delta(\cdot)$ is the determinant and $\bPhi=\sum_{i=1}^r {\mathbf{a}_i}{ \mathbf{a}_i^{\mathsf T}}$, where $\mathbf{a}_i \sim \cN(\mathbf{0}^n, \I_{n \times n})$. To achieve~\eqnref{pdf}, we use the chain rule to break the computation to smaller part. We then compute the probability distribution function of every term in the chain rule. 
Note that~\eqnref{pdf} is not a Wishart distribution as $\bA^{-1} \Phi$ need not be symmetric. This is the most technical part of the proof and we perform this computation from the basics; rest of the proof requires clever manipulation of the $\PDF$ for neighbouring matrices. 
\begin{extended}
We give a detail proof for $\mathsf{PSG}_1$ in~\appref{dp1} and $\mathsf{PSG}_2$ in~\appref{wishart}. 
\end{extended}

{\em $\mathsf{PSG}_2$ is interesting in its own regard. It says that two successive applications of a Gaussian matrix in a defined form preserves differential privacy}. In general, reuse of randomness does not preserve privacy, but what we show here is that if the randomness is reused cleverly, then it is possible to achieve privacy. 
\begin{full}
\begin{proof} 
We start with the proof of part (i).
We give the proof~\cite{BBDS12} for the sake of completion. We denote by $\widetilde{\bA}$ the matrix that differs from $\bA$ by at most one entry. In other word, if $\bA$ and $\widetilde{\bA}$ 
differs in row $i$, then there exists a unit vector $v$ such that $\bA - \widetilde{\bA}= \mathbf{E}= \mathbf{v} \mathbf{e}_i^{\mathsf T}  .$ Let $\bU \bSigma \bV^{\mathsf{T}}$ ($\widetilde{\bU} \widetilde{\bSigma} \widetilde{\bV}^{\mathsf{T}}$, respectively) be the $\mathsf{SVD}$ of $\bA$ ($\widetilde{\bA}$, respectively).

The {\sf PDF} for the two distributions, corresponding to $\bA$ and $\widetilde{\bA}$, is just a linear transformation of $\cN(0,\I_{n \times n})$. Therefore,
\begin{align*}
	\PDF_{\bA^{\mathsf T}Y} (\bx) & = \frac{1}{\sqrt{(2\pi)^d \Delta(\bA^{\mathsf T}  \bA)}} \exp(- \frac{1}{2} \bx (\bA^{\mathsf T}  \bA)^{-1} \bx^{\mathsf T})\\
	\PDF_{\widetilde{\bA}^{\mathsf T}Y} (\bx) & = \frac{1}{\sqrt{(2\pi)^d \Delta(\widetilde{\bA}^{\mathsf T}  \widetilde{\bA})}} \exp(- \frac{1}{2} \bx (\widetilde{\bA}^{\mathsf T}  \widetilde{\bA})^{-1} \bx^{\mathsf T}) 
\end{align*}

We prove the result for a row of the published matrix; the theorem follows from Theorem~\ref{thm:DRV}. 
It is straightforward to see that combination of the following proves differential privacy for a row of published matrix:
\begin{align*}  \sqrt{\frac{\Delta(\bA^{\mathsf T}  \bA)}{\Delta(\widetilde{\bA}^{\mathsf T}  \widetilde{\bA})}} \in  \exp(\pm \varepsilon_0) \quad \text{and} \quad 
\p \sparen{ \left| \bx   \bOmega^{\mathsf T}  (\bA^{\mathsf T}  \bA)^{-1}\bOmega\bx^{\mathsf T}- \bx  \bOmega^{\mathsf T}  (\widetilde{\bA}^{\mathsf T}   \widetilde{\bA})^{-1}\bOmega \bx^{\mathsf T} \right|   \leq \varepsilon_0} \geq 1 -\delta_0,\end{align*} where  $\varepsilon_0 = \frac{\varepsilon}{\sqrt{4 r \ln (2/\delta)}}$ and  $\delta_0 = {\delta}/{2r}. $

The first part of the proof follows simply as in~\cite{BBDS12}. More concretely, we have $\det(\bA^{\mathsf T}  \bA) = \prod_i \sigma_i^2$, where $\sigma_1 \geq \cdots \geq \sigma_d \geq  \sigma_{\mathsf{min}}$ are the singular values of $\bA$. 
 Let $\widetilde{\sigma}_1 \geq \cdots \geq \widetilde{\sigma}_d \geq  \sigma_{\mathsf{min}}$ be its singular value  for $\widetilde{\bA}$. Since the singular values of $\bA - \widetilde{\bA}$ and $\widetilde{\bA} -\bA$ are the same,  $\sum_i(\sigma_i - \widetilde{\sigma}_i) \leq 1$ using Linskii's theorem. Therefore, 
\begin{align*} \sqrt{\prod_i \frac{\widetilde{\sigma}_i^2}{\sigma_i^2}} \leq \exp \paren{\frac{\varepsilon}{32 \sqrt{r \log (2/\delta)} \log (r/\delta)} }\sum_i (\widetilde{\sigma}_i - \sigma_i) \leq e^{\varepsilon_0/2}. \end{align*}

The second part of the proof is slightly more involved.  
 Each row $i$ of the published matrix is distributed identically and is constructed by multiplying  an $n$-dimensional  vector $\bOmega_i$ that has entries 
picked 
from a normal distribution $\cN(0,1)$. Note that $\E[\bOmega_{i:}]=\mathbf{0}^n$ and $\cov(\bOmega_{i:})=\I$. Then
\begin{align*}	\bx   \bOmega^{\mathsf T}  (\bA^{\mathsf T}  \bA)^{-1}\bOmega \bx^{\mathsf T} - \bx   \bOmega^{\mathsf T}  (\widetilde{\bA}^{\mathsf T}   \widetilde{\bA})^{-1}\bOmega \bx^{\mathsf T} ={\bx   \bOmega^{\mathsf T}   \sparen{(\bA^{\mathsf T}  \bA)^{-1} (\bA^{\mathsf T}   \mathbf{E} +  \mathbf{E}^{\mathsf T}  \widetilde{\bA}) (\widetilde{\bA}^{\mathsf T}  
\widetilde{\bA})^{-1} } \bOmega  \bx^{\mathsf T}}.  \end{align*}

Using the singular value decomposition of $\bA=\bU \bSigma \bU^{\mathsf T}  $ and $\widetilde{\bA} = \widetilde{\bU} \widetilde{\bSigma} \widetilde{\bU}^{\mathsf T}  $, this simplifies as
\begin{align*} \paren{ \bx   \bOmega^{\mathsf T}  (\bV \bSigma^{-1}\bU^{\mathsf T}  ) \mathbf{e}_i} \paren{\mathbf{v}^{\mathsf T}   (\widetilde{\bV} \widetilde{\bSigma}^{-2} \widetilde{\bV}^{\mathsf T}  ) \bOmega \bx^{\mathsf T}} 
+    \paren{\bx   \bOmega^{\mathsf T} (\bV \bSigma^{-2}\bV^{\mathsf T}  )\mathbf{v}} \paren{\mathbf{e}_i^{\mathsf T}   (\widetilde{\bU} \widetilde{\bSigma}^{-1} \widetilde{\bV}^{\mathsf T}  ) \bOmega \bx^{\mathsf T}}.
\end{align*}

Since $\bx \sim \bA^{\mathsf T}   \by$, where $\by \sim \cN(0,1)$, we can further simplify it as 
				\begin{align*} \underbrace{\paren{ \by^{\mathsf T}  \bA \bOmega^{\mathsf T}  (\bV \bSigma^{-1}\bU^{\mathsf T}  ) \mathbf{e}_i}}_{t_1} \underbrace{\paren{\mathbf{v}^{\mathsf T}   (\widetilde{\bV} \widetilde{\bSigma}^{-2} \widetilde{\bV}^{\mathsf T}  ) \bOmega \bA^{\mathsf T}  \by} }_{t_2}
				+ \underbrace{\paren{\by^{\mathsf T}  A\bOmega^{\mathsf T}  (\bV \bSigma^{-2}\bV^{\mathsf T}  )\mathbf{v}}}_{t_3} \underbrace{\paren{\mathbf{e}_i^{\mathsf T}   (\widetilde{\bU} \widetilde{\bSigma}^{-1} \widetilde{\bV}^{\mathsf T}  ) \bOmega \bA^{\mathsf T}  \by}}_{t_4}. \end{align*}

Now since $\| \widetilde{\bSigma} \|_2, \| \bSigma \|_2 \geq w$, plugging in the $\mathsf{SVD}$ of $\bA$ and $\bA - \bA' = \mathbf{e}_i \mathbf{v}^{\mathsf T} $, and that every term $t_i$ in the above expression is a linear combination of a Gaussian, i.e., each term is distributed as per $ \cN(0,\|t_i\|^2)$, we calculate $\|t_i\|$ as below.
\begin{align*}
& \| (\bU \bSigma \bV^{\mathsf T} )  \bOmega^{\mathsf T}  (\bV \bSigma^{-1}\bU^{\mathsf T}  ) \mathbf{e}_i \|_2  \leq \| \bOmega^{\mathsf T}   \|_2 \leq 1, \\
& \| (\bU \bSigma \bV^{\mathsf T} ) \bOmega^{\mathsf T} (\bV \bSigma^{-2}\bV^{\mathsf T}  )\mathbf{v} \|_2   \leq \| \bOmega^{\mathsf T}   \|_2 \| \bSigma^{-1} \|_2 
\leq \frac{ 1}{ \sigma_{\mathsf{min}}},  \\
& \| v^{\mathsf T}   (\widetilde{\bV} \widetilde{\bSigma}^{-2} \widetilde{\bV}^{\mathsf T}  ) \bOmega (\widetilde{\bV} \widetilde{\bSigma} \widetilde{\bU}^{\mathsf T}   - \mathbf{v}\mathbf{e}_i^{\mathsf T}  )\|_2   \leq \| v^{\mathsf T}   (\widetilde{\bV} \widetilde{\bSigma}^{-2} \widetilde{\bV}^{\mathsf T}  ) \bOmega  \widetilde{\bV} \widetilde{\bSigma} 
\widetilde{\bU}^{\mathsf T}   
\|_2 + \| v^{\mathsf T}   (\widetilde{\bV} \widetilde{\bSigma}^{-2} \widetilde{\bV}^{\mathsf T}  ) \bOmega  \mathbf{v}\mathbf{e}_i^{\mathsf T}   \|_2  \\
& \qquad \leq \frac{1}{ \sigma_{\mathsf{min}}} + \frac{1 }{ \sigma_{\mathsf{min}}^2},  \\
& \| e_i^{\mathsf T}   (\widetilde{\bU} \widetilde{\bSigma}^{-1} \widetilde{\bV}^{\mathsf T}  ) \bOmega (\widetilde{\bV} \widetilde{\bSigma} \widetilde{\bU}^{\mathsf T}   - \mathbf{v}\mathbf{e}_i^{\mathsf T}  ) \|_2   \leq \|  \mathbf{e}_i^{\mathsf T}   (\widetilde{\bU} \widetilde{\bSigma}^{-1} \widetilde{\bV}^{\mathsf T}  ) \bOmega (\widetilde{\bV} \widetilde{\bSigma} 
\widetilde{\bU}^{\mathsf T}   \|_2 + \|\mathbf{e}_i^{\mathsf T}   (\widetilde{\bU} \widetilde{\bSigma}^{-1} \widetilde{\bV}^{\mathsf T}  ) \bOmega \mathbf{v}\mathbf{e}_i^{\mathsf T}   \|_2 \\ 
& \qquad \leq 1  + \frac{1 }{ \sigma_{\mathsf{min}}}, \end{align*}
 where $ \sigma_{\mathsf{min}}=\paren{ \frac{ \sqrt{r\log (2/\delta)} \log (r/ \delta)}{\varepsilon}}$.
 Using the concentration bound on the Gaussian distribution, each term, $t_1,t_2,t_3$, and $t_4$, is less than $\|t_i\| \ln (4/\delta_0)$ with probability $1 - \delta_0/2$. From the fact that $2\paren{\frac{1}{ \sigma_{\mathsf{min}}} + \frac{1}{ \sigma_{\mathsf{min}}^2}} \ln (4/\delta_0) \leq \varepsilon_0$, we have
the second part of the proof.

\medskip
We now prove that the second variant preserves privacy if the singular values of the streamed matrix follows the hypothesis of the theorem. 
We first compute the $\PDF$ of the published matrix. 

 We start by computing the probability density function when the underlying multivariate Gaussian distribution is $\cN(0,\I)$. The case for arbitrary positive definite covariance matrix follows  like the transition from identity to arbitrary positive definite covariance matrices in the multivariate Gaussian distribution. Let $\mathbf{g}_1, \cdots, \mathbf{g}_r\sim \cN(0,\I)$ be $r$ i.i.d multivariate Gaussian distribution, i.e., $\mathbf{g}_{ij} \sim \cN(0,1)$ for $1 \leq i \leq r, 1 \leq j \leq n$. The distribution we are interested in is $\bPhi=\sum_{i=1}^r \mathbf{g}_i \mathbf{g}_i^{\mathsf T}$. We use the notation $\PDF(\bPhi; \I)$ to denote the probability density function of $\bPhi$ when each random variable is picked using a normal distribution, i.e., when the covariance matrix of the random variables is $\I$.

Using the chain rule, the joint distribution of the entries of $\bPhi$ is  as follows.
\begin{align} 
\PDF(\bPhi; \I) &= \PDF(\brak{\mathbf{g}_1, \mathbf{g}_1}; \I) \PDF((\brak{\mathbf{g}_2, \mathbf{g}_1}, \brak{\mathbf{g}_2, \mathbf{g}_2} )|\brak{\mathbf{g}_1, \mathbf{g}_1} ; \I) \cdots \nonumber \\ & \qquad \PDF( (\brak{\mathbf{g}_r, \mathbf{g}_1}, \cdots \brak{\mathbf{g}_r, \mathbf{g}_r}) | \bPhi_{[r-1]} ; \I). \label{eq:pdf}
\end{align}
There are $r(r+1)/2$ distinct entries, $\brak{\mathbf{g}_1, \mathbf{g}_1}, ( \brak{\mathbf{g}_2, \mathbf{g}_1}, \brak{\mathbf{g}_2, \mathbf{g}_2}), \cdots, (\brak{\mathbf{g}_r, \mathbf{g}_1}, \cdots \brak{\mathbf{g}_r, \mathbf{g}_r})$.  Our aim is to compute each individual term in the product form of the above chain rule. For this, we  first analyze and understand the properties of the distribution of $\mathbf{h}_{i-1}^{\mathsf{T}} =(\brak{ \mathbf{g}_{i},\mathbf{g}_{1}}, \cdots , \brak{ \mathbf{g}_{i}, \mathbf{g}_{i-1}})$.  Then we use the fact that there is a transformation of Jacobian one from   
$ \paren{  \brak{ \mathbf{g}_{i}, \mathbf{g}_{1}} , \cdots , \brak{\mathbf{g}_{i}, \mathbf{g}_{(i-1)}} , { \brak{ \mathbf{g}_{i}, \mathbf{g}_{i}} - \mathbf{h}_{i-1}^{\mathsf{T}} \bPhi_{[i-1]}^{-1} \mathbf{h}_{i-1} }}$ to $\paren{  \brak{ \mathbf{g}_{i}, \mathbf{g}_{1}} , \cdots , \brak{\mathbf{g}_{i}, \mathbf{g}_{(i)}} }$ to compute each term in the chain rule~(see, \cite{Muirhead}~and~\cite{Rao73}).

\paragraph{Distribution of $\mathbf{h}_{i-1}$.} 
We first prove that $\mathbf{h}_{i-1}$ is an $(i-1)$-variate Gaussian distribution. 
Since the covariance matrix is $\I,$ and $\mathbf{g}_{11}, \cdots,  \mathbf{g}_{1n}, \cdots, \mathbf{g}_{r1}, \cdots, \mathbf{g}_{rn}$ are i.i.d. $\cN(0,1)$, from the elementary property of linear functions of normal variables, conditional on $\mathbf{g}_{kj}$ for $1 \leq k \leq i-1$ and $1 \leq j \leq n$, $\mathbf{h}_{i-1}$ is $(i-1)$-variate Gaussian distribution with 
\begin{align*} \bPhi_{[i]} =\paren{ \begin{matrix}  \brak{\mathbf{g}_{1}, \mathbf{g}_{1}} & \cdots & \brak{ \mathbf{g}_{1}, \mathbf{g}_{i}} \\
					\vdots & \ddots &  \vdots \\
					 \brak{ \mathbf{g}_{i}, \mathbf{g}_{1}} &  \cdots &  \brak{\mathbf{g}_{i}, \mathbf{g}_{i}}   
		\end{matrix} } \end{align*}

Now $\mathbf{g}_{11}, \cdots, \mathbf{g}_{rn}$, for every $j=1, \cdots ,n$ are mutually independent; therefore, we have the covariance 
\begin{align*}  \cov (\mathbf{h}_{i-1},\mathbf{g}_{ij}) = \sparen{ \cov \paren{ \brak{ \mathbf{g}_{i}, \mathbf{g}_{1}} \mathbf{g}_{ij} }, \cdots , \cov \paren{ \brak{ \mathbf{g}_{i}, \mathbf{g}_{(i-1)}} \mathbf{g}_{ij}}  }  =  (\mathbf{g}_{1j}, \cdots , \mathbf{g}_{i-1,j})^{\mathsf{T}} \end{align*} 
and $\E [ \ket{\mathbf{h}_{i-1}} \bra{ \mathbf{h}_{i-1}} | \mathbf{g}_{kj}] = \bPhi_{[i-1]}$ for $1 \leq j <i$. This implies
\begin{align}  \cov \sparen{ \mathbf{h}_{i-1}, \mathbf{g}_{ij}- \mathbf{h}_{i-1}^{\mathsf{T}} \bPhi^{-1}_{[i-1]} (\mathbf{g}_1j , \cdots \mathbf{g}_{i-1,j})^{\mathsf{T}} | \mathbf{g}_{kj} } = 0 \qquad \forall 1 \leq k \leq i-1, \label{eq:covariance} \end{align}
as the left hand side equals $ (\mathbf{g}_1j , \cdots \mathbf{g}_{i-1,j})^{\mathsf{T}}  - \bPhi_{[i-1]} \bPhi^{-1}_{[i-1]}  (\mathbf{g}_1j , \cdots \mathbf{g}_{i-1,j})^{\mathsf{T}} . $

This implies that $\mathbf{h}_{i-1}$ is independent of $\sum_{j=1}^k \paren{\mathbf{g}_{ij} - (\mathbf{g}_{1j} , \cdots , \mathbf{g}_{i-1,j} )^{\mathsf{T}} \bPhi_{[i-1]}^{-1} (\mathbf{g}_{1j , \cdots , \mathbf{g}_{i-1,j} }) }^2$.~\cite{Rao73} proved that  
\begin{align} \sum_{j=1}^k \paren{\mathbf{g}_{ij} - (\mathbf{g}_{1j} , \cdots , \mathbf{g}_{i-1,j} )^{\mathsf{T}} \bPhi_{[i-1]}^{-1} (\mathbf{g}_{1j} , \cdots , \mathbf{g}_{i-1,j} ) }^2 \sim \chi^2_{n-i+1},\label{eq:chi}  \end{align} the standard $\chi^2$ distribution with $(n-i+1)$ degrees of freedom.

\paragraph{Computing every term in the chain rule.} From the fact that $\mathbf{h}_{i-1}$ is a $(i-1)$-variate Gaussian distribution,~\eqnref{covariance},~\eqnref{chi}, and the identity 
\begin{align*} \Delta(\bPhi_{[i]}) = \Delta(\bPhi_{[i-1]}) \sum_{j=1}^k \paren{\mathbf{g}_{ij} - (\mathbf{g}_{1j} , \cdots , \mathbf{g}_{i-1,j} )^{\mathsf{T}} \bPhi_{[i-1]}^{-1} (\mathbf{g}_{1j} , \cdots , \mathbf{g}_{i-1,j} ) }^2 ,\end{align*} 
where $\Delta(\cdot)$ denotes the determinant, we first calculate the joint pdf of 
\begin{align}
	& \paren{  \brak{ \mathbf{g}_{i}, \mathbf{g}_{1}} , \cdots , \brak{\mathbf{g}_{i}, \mathbf{g}_{(i-1)}} , \brak{ \mathbf{g}_{i}, \mathbf{g}_{i}} - {\mathbf{h}^{\mathsf T}_{i-1}} \bPhi^{-1}_{[i-1]} {\mathbf{h}_{i-1}} }^{\mathsf{T}} \nonumber \\
	& \qquad = \frac{ \exp \paren{ -\frac{1}{2}  \paren{  \brak{ \mathbf{g}_{i}, \mathbf{g}_{1}} , \cdots , \brak{\mathbf{g}_{i}, \mathbf{g}_{(i-1)}} }^{\mathsf{T}} \bPhi^{-1}_{[i-1]}  \paren{ \brak{ \mathbf{g}_{i}, \mathbf{g}_{1}} , \cdots , \brak{ \mathbf{g}_{i} \mathbf{g}_{(i-1)}}} } }{ (2 \pi)^{(i-1)/2} \Delta(\bPhi_{[i-1})^{1/2}} \nonumber  \\
	& \qquad \qquad \times \frac{ \exp \paren{  -\frac{ \brak{ \mathbf{g}_{i}, \mathbf{g}_{i}} -  \mathbf{h}_{i-1}^{\mathsf{T}} \bPhi^{-1}_{[i-1]} \mathbf{h}_{i-1}}{2} } \paren{ \brak{ \mathbf{g}_{i}, \mathbf{g}_{i}} - \mathbf{h}_{i-1}^{\mathsf{T}} \bPhi_{[i-1]}^{-1} \mathbf{h}_{i-1} }^{(n-i+1)/2 -1} }{ 2^{(n-i+1)/2} \Gamma ((n-i+1)/2) } \nonumber  \\
	& \qquad = \frac{ \exp \paren{ - \frac{\mathbf{h}_{i-1}^{\mathsf{T}} \bPhi_{[i-1]}^{-1} \mathbf{h}_{i-1} }{2} } \Delta(\bPhi_i)^{(n-i-1)/2} }{ 2^{m/2} \pi^{(i-1)/2} \Gamma((n-i+1)/2) \Delta(\bPhi_{[i-1]})^{(n-i)/2}}  \nonumber \\
	& \qquad =  \PDF ((\brak{\mathbf{g}_i, \mathbf{g}_1}, \cdots , \brak{\mathbf{g}_i, \mathbf{g}_{i-1}}, \brak{\mathbf{g}_i, \mathbf{g}_i} ) | \bPhi_{[i-1]}),\label{eq:4}
\end{align}
where the last step uses the fact that the transformation  from 
$ \paren{  \mathbf{h}_i^{\mathsf T} , { \brak{ \mathbf{g}_{i}, \mathbf{g}_{i}} - \mathbf{h}_{i-1}^{\mathsf{T}} \bPhi_{[i-1]}^{-1} \mathbf{h}_{i-1} }}$ to \\ $\paren{  \brak{ \mathbf{g}_{i}, \mathbf{g}_{1}} , \cdots , \brak{\mathbf{g}_{i}, \mathbf{g}_{(i)}} }$ is one-to-one with Jacobian $1$.

\paragraph{Computing the joint distribution of $\bPhi$.} We are done except for putting in the values of every terms in the chain rule. A simple arithmetic followed by plugging~\eqnref{4} in~\eqnref{pdf} gives the closed formed expression of the pdf of $\bPhi$ as
\begin{align*} 
\frac{ \exp (-\tr (\bPhi)/2) \Delta(\bPhi)^{(n-r-1)/2}} {2^{rn/2} \pi^{\sum_i (i-1)/2} \prod_{i=1}^r \Gamma((n-i+1)/2)} & \times \prod_{i=1}^r \paren{ \frac{\Delta(\bPhi_{[i]})^{(n-i-1)/2}} {\Delta(\bPhi_{[i-1]})^{(n-i)/2}} } \\
&= 
\frac{ \exp (-\tr (\bPhi)/2) \Delta(\bPhi)^{(n-r-1)/2}} {2^{rn/2} \pi^{n(n-1)/4} \prod_{i=1}^r \Gamma((n-i+1)/2)}. 
\end{align*}

Let $\bPhi$ be the matrix formed in the manner as above with random vectors picked using the distribution that is defined by $\cN(\mathbf{0}^n, \bSigma)$. Let $\bPhi'= \bPhi \bA$. Then we can write the probability distribution function of $\bPhi'$ as follows
\begin{align*}
\PDF_{\bPhi'}(\bPhi') & \propto \PDF_\bPhi (\bPhi' \bA^{-1} ) \\
& \propto \exp \paren{ \tr\paren{- \bSigma^{-1} \bPhi \bA^{-1}/2} }\\
&=  \exp\paren{ \tr \paren{-\bA^{-1} \bSigma^{-1} \bPhi }/2 },
\end{align*}
where $\bA^{-1}$ is the pseudo-inverse of $\bA$. This is not distributed like the distribution of $\bPhi$.

We can now prove the privacy guarantee. Let $\delta_0= \delta/r$.
Let $\bA$ and $\widetilde{\bA}$ be a pair of neighbouring matrices that form the stream. From our definition of neighbouring matrices, $ \bA -\widetilde{\bA}   = \mathbf{E} = \ket{\mathbf{v}}  \bra{\mathbf{e}_i}$ for an unit vector $\ket{\mathbf{v}}$ and some $i$. The published matrices corresponding to the two neighbouring matrices have the following probability density function
\begin{align*}
	\PDF(\bPhi \bA; \I ) &=   \frac{ \exp (-\tr (\bA^{-1} \bPhi )/2) \Delta(\bPhi)^{(n-r-1)/2}} {2^{rn/2} \pi^{n(n-1)/4} \Delta(\bA)^{r/2} \prod_{i=1}^r \Gamma((n-i+1)/2)} = C \frac{ \exp (-\tr (\bA^{-1} \bPhi )/2) } { \Delta(\bA)^{r/2} }, \\
	\PDF( \bPhi \widetilde{\bA} ; \I ) &=   \frac{ \exp (-\tr ((\widetilde{\bA}^{-1} \bPhi )/2) \Delta( \bPhi)^{(n-r-1)/2}} {2^{rn/2} \pi^{n(n-1)/4} \Delta(\widetilde{\bA} )^{r/2} \prod_{i=1}^r \Gamma((n-i+1)/2)} = C  \frac{ \exp (-\tr ((\widetilde{\bA}^{\mathsf{T}} \widetilde{\bA})^{-1} \bPhi )/2)} { \Delta(\widetilde{\bA} )^{r/2}}, 
\end{align*}
where $ C= { \Delta( \bPhi)^{(n-r-1)/2}}/ \paren{2^{rn/2} \pi^{n(n-1)/4} \prod_{i=1}^r \Gamma((n-i+1)/2)}$.
It is straightforward to see that combination of the following proves differential privacy of the published matrix:
\begin{align} \exp(-\varepsilon/r) \leq \sqrt{\frac{\widetilde{\Delta}(\bA)}{\widetilde{\Delta}(\widetilde{\bA})}} \leq  \exp( \varepsilon/r) \quad \text{and} \quad 
\p \sparen{| \tr \paren{ \paren{ \bA^{-1} - \widetilde{\bA}^{-1} } \bPhi} |   \leq \varepsilon} \geq 1 -\delta. \label{eq:main}
\end{align} 

Let $\sigma_1\geq \cdots, \geq \sigma_d \geq \sigma_{\mathsf{min}}$ be the singular values of $\bA$. 
 Let $\widetilde{\sigma}_1, \geq \cdots, \geq \widetilde{\sigma}_d \geq \sigma_{\mathsf{min}}$ be the singular value  for $\widetilde{\bA}$. Since the singular values of $\bA - \widetilde{\bA}$ and $\widetilde{\bA} -\bA$ are the same,  $\sum_{i \in G}(\sigma_i - \widetilde{\sigma}_i) \leq 1$ using Linskii's theorem, where $G$ is the set of indices for which $\sigma_i > \widetilde{\sigma}_i$. The first bound follows similarly as in Blocki {\it et al.}~\cite{BBDS12}.
 For the second  bound required for the privacy, we first bound the following
\begin{align*}
 \tr \paren{ \paren{ \bA^{-1} - \widetilde{\bA}^{-1} } \bPhi}  &=  \tr \paren{ \paren{ \bA^{-1} (\widetilde{\bA}) (\widetilde{\bA})^{-1} - \widetilde{\bA}^{-1} } \bPhi} \\
 		&=  \tr \paren{ \paren{ \bA^{-1} (\bA+\mathbf{E}) \widetilde{\bA}^{-1} -  \widetilde{\bA}^{-1} } \bPhi} 
		=  \tr \paren{ \paren{ \bA^{-1} \mathbf{E} \widetilde{\bA}^{-1} } \bPhi }.
 \end{align*}

Using the singular value decomposition of $\bA = \bU \bSigma \bV^{\mathsf{T}}$ and $\widetilde{\bA} = \widetilde{\bU} \widetilde{\bSigma} \widetilde{\bV}^{\mathsf{T}}$, and the fact that $\mathbf{E} = {\mathbf{v}} { \mathbf{e}_i}^{\mathsf T}$ for some $i$, we can further solve the above expression.
\begin{align*}
  \tr \paren{ \paren{ \bA^{-1} - \widetilde{\bA}^{-1} } \bPhi}  &= \tr \paren{ \paren{ \bA^{-1} \mathbf{E} \widetilde{\bA}^{-1} } \bPhi }  \\
  	&= \tr \paren{  \bV \bSigma^{-1} \bU^{\mathsf{T}} {\mathbf{v}} { \mathbf{e}_i}^{\mathsf T} \widetilde{\bU} \widetilde{\bSigma}^{-1} \widetilde{\bV}^{\mathsf{T}}\bPhi }   \\
	&= \sum_{j=1}^r \tr  \paren{ \mathbf{g}^{\mathsf T}_j \bV \bSigma^{-1} \bU^{\mathsf{T}}  {\mathbf{v}} { \mathbf{e}_i}^{\mathsf T} \widetilde{\bU} \widetilde{\bSigma}^{-1} \widetilde{\bV}^{\mathsf{T}}\mathbf{g}_j }.
  \end{align*}

Fix a $j$. We bound the following.
\begin{equation}
 \left| \tr  \paren{ \mathbf{g}^{\mathsf T}_j \bV \bSigma^{-1} \bU^{\mathsf{T}}  {\mathbf{v}} { \mathbf{e}_i^{\mathsf T}} \widetilde{\bU} \widetilde{\bSigma}^{-1} \widetilde{\bV}^{\mathsf{T}}\mathbf{g}_j } \right| .\label{eqn:final}
\end{equation}

We now look at each term in the above expression. $ \mathbf{g}^{\mathsf T}_j \bV \bSigma^{-1} \bU^{\mathsf{T}} {\mathbf{e}_i}$ as $\cN(0, \|  \bV \bSigma^{-1} \bU^{\mathsf{T}} {\mathbf{e}_i}\|^2)$, 
and ${ \mathbf{v}^{\mathsf T}} \widetilde{\bU} \widetilde{\bSigma}^{-1} \widetilde{\bV}^{\mathsf{T}}\mathbf{g}_j$ as $\cN(0, \| { \mathbf{v}^{\mathsf T}} \widetilde{\bU} \widetilde{\bSigma}^{-1} \widetilde{\bV}^{\mathsf{T}} \|^2)$. Since $\mathbf{v}$ and $\mathbf{e}_i$ are unit vectors, the norm of the above four quantities are less than $ 1/\sigma_{\mathsf{min}},$ and $1+1/ \sigma_{\mathsf{min}}$, respectively. 

Therefore, from the concentration inequality of Gaussian distribution, we have 
\begin{align*} \p \sparen{ (\ref{eqn:final}) \leq 2 \paren{\frac{1}{ \sigma_{\mathsf{min}}} + \frac{1}{ \sigma_{\mathsf{min}}^2}} \ln (4/\delta_0) \leq \varepsilon } \geq 1- \delta_0. \end{align*}

Taking union bound, we have  with probability $1-\delta$, $-\varepsilon \leq \tr \paren{ \paren{ \bA^{-1} - \widetilde{\bA}^{-1} } \bPhi} \leq \varepsilon$.

\Jcom{
Using standard techniques (that could be found in any standard textbook, including~\cite{Rao73}) of the transformation method and the factorization theorem yields the $\PDF$ for arbitrary linear translation. That is, if $X \sim \PDF(\bPhi; \I)$, then $Y=AX$ is distributed as $\PDF(\bPhi; \bA^{\mathsf{T}}  A)$.  The proof is similar to the similar transformation for multivariate Gaussian distribution. It is easy to verify that it does not matter if we multiply $\bA$ from right or left of vectors $\mathbf{g}_1, \cdots, \mathbf{g}_r$, i.e., $\sum_{i=1}^r A \mathbf{g}_i \mathbf{g}_i^{\mathsf T}$ and $ \sum_{i=1}^r \mathbf{g}_i \mathbf{g}_i^{\mathsf T}\bA^{\mathsf T}$ have the same distribution. More concretely, for $\sum_{i=1}^r A \mathbf{g}_i \mathbf{g}_i^{\mathsf T}$, the distribution is
\begin{align*} 
\PDF (\bPhi ; \bA^{\mathsf T}A) &= \PDF (\bPhi; \I) \frac{\PDF(\brak{\mathbf{g}_1, \cdots, \mathbf{g}_r} ; \bA^{\mathsf T}A) } {\PDF(\brak{\mathbf{g}_1, \cdots, \mathbf{g}_r} ; \I)} \\
	 &= \frac{ \exp (-\tr ((\bA^{\mathsf{T}}A)^{-1} \bPhi )/2) \Delta(\bPhi)^{(n-r-1)/2}} {2^{rn/2} \pi^{r(r-1)/4} \Delta(\bA^{\mathsf{T}} A)^{r/2} \prod_{i=1}^r \Gamma((n-i+1)/2)}. \end{align*}
}

\Jcom
{We can now prove the privacy guarantee. Let $\delta_0= \delta/r$.
Let $\bA$ and $\widetilde{\bA}$ be a pair of neighbouring matrices that form the stream. From our definition of neighbouring matrices, $ A -\widetilde{\bA}   = E = \ket{\mathbf{v}}  \bra{\mathbf{e}_i}$ for an unit vector $\ket{\mathbf{v}}$ and some $i$. The published matrices corresponding to the two neighbouring matrices have the following probability density function
\begin{align*}
	\PDF(\bPhi A; \I ) &=   \frac{ \exp (-\tr (\bA^{-1} \bPhi )/2) \Delta(\bPhi)^{(n-r-1)/2}} {2^{rn/2} \pi^{n(n-1)/4} \Delta(\bA)^{r/2} \prod_{i=1}^r \Gamma((n-i+1)/2)} = C \frac{ \exp (-\tr (\bA^{-1} \bPhi )/2) } { \Delta(\bA)^{r/2} }, \\
	\PDF( \bPhi \widetilde{\bA} ; \I ) &=   \frac{ \exp (-\tr ((\widetilde{\bA}^{-1} \bPhi )/2) \Delta( \bPhi)^{(n-r-1)/2}} {2^{rn/2} \pi^{n(n-1)/4} \Delta(\widetilde{\bA} )^{r/2} \prod_{i=1}^r \Gamma((n-i+1)/2)} = C  \frac{ \exp (-\tr ((\widetilde{\bA}^{\mathsf{T}} \widetilde{\bA})^{-1} \bPhi )/2)} { \Delta(\widetilde{\bA} )^{r/2}}, 
\end{align*}
where $ C= { \Delta( \bPhi)^{(n-r-1)/2}}/ \paren{2^{rn/2} \pi^{n(n-1)/4} \prod_{i=1}^r \Gamma((n-i+1)/2)}$.
It is straightforward to see that combination of the following proves differential privacy of the published matrix:
\begin{align} \exp(-\varepsilon/r) \leq \sqrt{\frac{\widetilde{\Delta}(\bA^{\mathsf T}  A)}{\widetilde{\Delta}(\bar{\bA}^{\mathsf T}  \bar{\bA})}} \leq  \exp( \varepsilon/r) \quad \text{and} \quad 
\p \sparen{| \tr \paren{ \paren{ (\bA^{\mathsf{T}}A)^{-1} - (\widetilde{\bA}^{\mathsf{T}} \widetilde{\bA})^{-1} } \bPhi} |   \leq \varepsilon} \geq 1 -\delta. \label{eq:main}
\end{align} 

Let $\sigma_1\geq \cdots, \geq \sigma_d \geq \sigma_{\mathsf{min}}$ be the singular values of $\bA$. 
 Let $\widetilde{\sigma}_1, \geq \cdots, \geq \widetilde{\sigma}_d \geq \sigma_{\mathsf{min}}$ be the singular value  for $\widetilde{\bA}$. Since the singular values of $A - \widetilde{\bA}$ and $\widetilde{\bA} -\bA$ are the same,  $\sum_{i \in G}(\sigma_i - \widetilde{\sigma}_i) \leq 1$ using Linskii's theorem, where $G$ is the set of indices for which $\sigma_i > \widetilde{\sigma}_i$. The first bound follows similarly as in~\cite{BBDS12}.
 For the second  bound required for the privacy, we first bound the following
\begin{align*}
 \tr \paren{ \paren{ (\bA^{\mathsf{T}}A)^{-1} - (\widetilde{\bA}^{\mathsf{T}} \widetilde{\bA})^{-1} } \bPhi}  &=  \tr \paren{ \paren{ (\bA^{\mathsf{T}}A)^{-1} (\widetilde{\bA}^{\mathsf{T}} \widetilde{\bA}) (\widetilde{\bA}^{\mathsf{T}} \widetilde{\bA})^{-1} - (\widetilde{\bA}^{\mathsf{T}} \widetilde{\bA})^{-1} } \bPhi} \\
 		&=  \tr \paren{ \paren{ (\bA^{\mathsf{T}}A)^{-1} (A+E)^{\mathsf{T}} (A+E) (\widetilde{\bA}^{\mathsf{T}} \widetilde{\bA})^{-1} - (\widetilde{\bA}^{\mathsf{T}} \widetilde{\bA})^{-1} } \bPhi} \\
		&=  \tr \paren{ \paren{ (\bA^{\mathsf{T}} A)^{-1} (\bA^{\mathsf{T}} E + E^{\mathsf{T}} \widetilde{\bA}) (\widetilde{\bA}^{\mathsf{T}} \widetilde{\bA})^{-1} } \bPhi }.
 \end{align*}

Using the singular value decomposition of $A = U \bSigma \bV^{\mathsf{T}}$ and $\widetilde{\bA} = \widetilde{U} \widetilde{\bSigma} \widetilde{\bV}^{\mathsf{T}}$, and the fact that $E = \ket{\mathbf{v}} \bra{ \mathbf{e}_i}$ for some $i$, we can further solve the above expression.
\begin{align*}
  \tr \paren{ \paren{ (\bA^{\mathsf{T}}A)^{-1} - (\widetilde{\bA}^{\mathsf{T}} \widetilde{\bA})^{-1} } \bPhi}  &= \tr \paren{ \paren{ \bV \bSigma^{-1} U^{\mathsf{T}} \ket{\mathbf{e}_i} \mathbf{v}^{\mathsf T}  \widetilde{\bV} \widetilde{\bSigma}^{-2} \widetilde{\bV}^{\mathsf{T}} + \bV \bSigma^{-2} \bV^{\mathsf{T}} \ket{\mathbf{v}} \bra{ \mathbf{e}_i} \widetilde{U} \widetilde{\bSigma}^{-1} \widetilde{\bV}^{\mathsf{T}} } \bPhi}  \\
  	&= \tr \paren{  \bV \bSigma^{-1} U^{\mathsf{T}} \ket{\mathbf{e}_i} \mathbf{v}^{\mathsf T}  \widetilde{\bV} \widetilde{\bSigma}^{-2} \widetilde{\bV}^{\mathsf{T}} \bPhi } + \tr \paren{\bV \bSigma^{-2} \bV^{\mathsf{T}} \ket{\mathbf{v}} \bra{ \mathbf{e}_i} \widetilde{U} \widetilde{\bSigma}^{-1} \widetilde{\bV}^{\mathsf{T}}\bPhi }   \\
	&= \sum_{j=1}^r \tr  \paren{ \mathbf{g}^{\mathsf T}_j \bV \bSigma^{-1} U^{\mathsf{T}} \ket{\mathbf{e}_i} \mathbf{v}^{\mathsf T}  \widetilde{\bV} \widetilde{\bSigma}^{-2} \widetilde{\bV}^{\mathsf{T}} \mathbf{g}_j }\\ 
	& \qquad +  \sum_{j=1}^r \tr \paren{\mathbf{g}^{\mathsf T}_j\bV \bSigma^{-2} \bV^{\mathsf{T}} \ket{\mathbf{v}} \bra{ \mathbf{e}_i} \widetilde{U} \widetilde{\bSigma}^{-1} \widetilde{V}^{\mathsf{T}}\mathbf{g}_j }.
  \end{align*}

Fix a $j$. We bound the following.
\begin{equation}
 | \tr  \paren{ \mathbf{g}^{\mathsf T}_j V \bSigma^{-1} U^{\mathsf{T}} \ket{\mathbf{e}_i} \bra{\mathbf{v}}  \widetilde{V} \widetilde{\bSigma}^{-2} \widetilde{V}^{\mathsf{T}} \mathbf{g}_j } 
	 +  \tr \paren{\mathbf{g}^{\mathsf T}_jV \bSigma^{-2} V^{\mathsf{T}} \ket{\mathbf{v}} \bra{ \mathbf{e}_i} \widetilde{U} \widetilde{\bSigma}^{-1} \widetilde{V}^{\mathsf{T}}\mathbf{g}_j }| .\label{eqn:final}
\end{equation}

We now look at each term in the above expression. $\mathbf{g}^{\mathsf T}_jV \bSigma^{-2} V^{\mathsf{T}} \ket{\mathbf{v}}$ is distributed as $\cN(0, \| V \bSigma^{-2} V^{\mathsf{T}} \ket{\mathbf{v}} \|^2)$, $\bra{\mathbf{v}}  \widetilde{V} \widetilde{\bSigma}^{-2} \widetilde{V}^{\mathsf{T}} \mathbf{g}_j$ as $\cN(0, \|\bra{\mathbf{v}}  \widetilde{V} \widetilde{\bSigma}^{-2} \widetilde{V}^{\mathsf{T}} \|^2)$, $ \mathbf{g}^{\mathsf T}_j V \bSigma^{-1} U^{\mathsf{T}} \ket{\mathbf{e}_i}$ as $\cN(0, \|  V \bSigma^{-1} U^{\mathsf{T}} \ket{\mathbf{e}_i}\|^2)$, 
and $\bra{ \mathbf{e}_i} \widetilde{U} \widetilde{\bSigma}^{-1} \widetilde{V}^{\mathsf{T}}\mathbf{g}_j$ as $\cN(0, \| \bra{ \mathbf{e}_i} \widetilde{U} \widetilde{\bSigma}^{-1} \widetilde{V}^{\mathsf{T}} \|^2)$. Since $\mathbf{v}$ and $\mathbf{e}_i$ are unit vectors, the norm of the above four quantities are less than $1/ \sigma_{\mathsf{min}}^2, 1/ \sigma_{\mathsf{min}}^2+1/ \sigma_{\mathsf{min}}^3, 1/\sigma_{\mathsf{min}},$ and $1/ \sigma_{\mathsf{min}}+1/ \sigma_{\mathsf{min}}^2$, respectively. 

Therefore, from the concentration inequality of Gaussian distribution, we have 
\begin{align*} \p \sparen{ (\ref{eqn:final}) \leq 2 \paren{\frac{1}{ \sigma_{\mathsf{min}}^2} + \frac{1}{ \sigma_{\mathsf{min}}^3}} \ln (4/\delta_0) \leq \varepsilon } \geq 1- \delta_0. \end{align*}

Taking union bound, we have  with probability $1-\delta$, $-\varepsilon \leq \tr \paren{ \paren{ (\bA^{\mathsf{T}}A)^{-1} - (\widetilde{\bA}^{\mathsf{T}} \widetilde{\bA})^{-1} } \bPhi} \leq \varepsilon$.
}

\end{proof}
\end{full}

\paragraph{Differences between $\mathsf{PSG}_1$ and Blocki {\it et al.}~\cite{BBDS12}.} At high level, $\mathsf{PSG}_1$ has some resemblance to the mechanism of Blocki {\it et al.}~\cite{BBDS12} and Upadhyay~\cite{Upadhyay13} if we use  random Gaussian matrix as $\bOmega$. However, the analogy ends here, for eg., Blocki {\it et al.}~\cite{BBDS12} and Upadhyay~\cite{Upadhyay13} perform an affine transformation to convert the private matrix into a set of $\{\sqrt{w/n},1\}^n$ vectors, while we perform perturbation to raise the singular values before invoking $\mathsf{PSG}$ (see Sections~\ref{sec:lsi} and~\ref{sec:application}).
The mechanism of~\cite[Algorithm 3]{BBDS12} does not give a guarantee that the singular values of $\bA^{\mathsf T}\bA$ and their published matrix is close or their eigenvalues are comparable. In other words, it does not give a $\lra$. Apart from these major differences, there are couple of  subtle differences: (i) they project the entries of the columns of  private matrix to a higher dimensional space; here, we perform embedding to a lower dimensional subspace in the similar vein as other applications of dimensionality reduction, and (ii) their mechanism uses multiple passes over the input matrix (they require at least two-passes over the input matrix even with the streaming algorithms for computing the  singular value decomposition ($\mathsf{SVD}$). So, we cannot use their mechanism in any of the problems we study in this paper.

\section{Low Rank Approximation} \label{sec:lsi}
Blocki {\it et al.}~\cite{BBDS12} noted that their published matrix is neither close nor their eigenvalues are comparable to the private matrix. In other words, it does not give a $\lra$; therefore, we need a different approach. We use the prototype mentioned in Halko  {\it et al.}~\cite{HMT11}, which was also used by Hardt and Roth~\cite{HR12} to improve the worst-case bound under a {\em low coherence} assumption.  In this prototype, we construct a low-dimensional subspace that captures the action of the matrix ({\em range-finding}), and then restrict the matrix to that subspace to compute the required  factorization ({\em projection}). More concretely, range finding finds a measurement matrix $\bY= \bA \bOmega$, where $\bOmega$ is a Gaussian matrix in our case and computes the orthonormal projection matrix $\Pi_\bY$ corresponding to the range defined by $\bY$; projection then computes a  $k$-rank  matrix $\mathbf{B}=\Pi_\bY \bA$. From this exposition, it seems that privacy preserving algorithms are required for both the stages; however, we show that the two-step prototype can be replaced by a two-step algorithm in which the input matrix is explicitly needed only in the first step at the expense of privacy proof requiring both $\mathsf{PSG}_1$ and $\mathsf{PSG}_2$. 

We first note that if $\bPsi \bPsi^{\mathsf{T}} \bA$ is a $\lra$ of $\bA$, i.e., $\| \bA - \bPsi \bPsi^ T \bA \| \leq \eta$, then so is $\bPsi \bPsi^{\mathsf{T}} \bA \bPsi \bPsi^{\mathsf{T}}$. This is because $ \| \bA -  \bPsi \bPsi^{\mathsf{T}} \bA\bPsi \bPsi^{\mathsf{T}}\| = \| \bA - \bPsi \bPsi^{\mathsf{T}} \bA + \bPsi \bPsi^{\mathsf{T}} \bA - \bPsi \bPsi^{\mathsf{T}} \bA \bPsi \bPsi^{\mathsf{T}} \| \leq \| \bA - \bPsi \bPsi^{\mathsf{T}} \bA \|  + \| \bPsi \bPsi^{\mathsf{T}} \bA - \bPsi \bPsi^{\mathsf{T}} \bA \bPsi \bPsi^{\mathsf{T}} \|  \leq 2 \eta. $
The crucial observation now for the single pass $\lra$ when $\bOmega$ is a Gaussian matrix is that $\bOmega$, $\bY$, and the basis for the range of $\bY$ contains enough information to compute the matrix $\mathbf{B}$, i.e., we do not need $\bA$ explicitly. The range-finding is private using~\thmref{PSG}(i), but as we  reuse $\bOmega$, we have to rely on~\thmref{PSG}(ii) to prove the privacy of projection step. In order to simplify the presentation, we  state our mechanism for symmetric matrices in~\figref{lsi}~(Hardt and Roth~\cite{HR13}~and Kapralov and Talwar~\cite{KT13} also made this assumption) that computes a $\lra$ of the $\mathsf{SVD}$ of $\bA$. 
\begin{extended}
The case for non-symmetric matrices and a proof of~\thmref{lsi} is covered in~\appref{lra}. 
\end{extended}

The mechanism for $\lra$ presented in~\figref{lsi} assume that the matrix $\bA$ is provided as the symmetric $\mathsf{rank}(\bA)$ matrix $\bA'$ corresponding to $\bA$, and stop updating the data structure once all the rows of $\bA$ are streamed. This simplifies the presentation as well as the analysis. By the argument of~\cite[Fact 2.8]{HR12}, this leads to a depreciation of the privacy guarantee by half (both $\varepsilon$ and $\delta$).  The analysis for error bound though has complications because the right and the left singular vectors of the original matrix might be different. Keeping this in mind, we present its analysis in the most general form.

\begin{figure} [t]
\fbox{
\begin{minipage}[l]{6.3in}
\small
{
On input parameters $\alpha, \beta, \varepsilon, \delta$, the target rank $k$, set $w=  \paren{ck\varepsilon^{-1} \ln(k/\delta)}$ for a global constant $c$. Pick a $2n \times k$ standard Gaussian matrix $\bOmega$. On input an $n \times n$ matrix $\bA$ of rank $r$, the mechanism does the following:

\begin{description}
	\item [Range Finding.] Compute $\bY_\bA = \begin{pmatrix} w \I & A \end{pmatrix} \bOmega$ by computing $\begin{pmatrix} w \mathbf{e}_i & \bA_{i:} \end{pmatrix} \bOmega$ for all streams $\bA_{i:},1 \leq i \leq n$ and appending them row-wise. Let $\Pi_{\bY_\bA} = \bPsi \bPsi^{\mathsf T} $ be the projection matrix corresponding to the range of $\bY_\bA$.
	\item [Projection:] When the whole matrix is streamed, the curator does the following:
\begin{enumerate}
	\item Let the (unknown) matrix $\mathbf{B}=\bPsi ^{\mathsf T}\bA_t \bPsi  $. Use the minimal residual method to find a solution to $\mathbf{B} \bPsi_t^{\mathsf T} \bOmega = \bPsi_t^{\mathsf T}  \bY_t$.
	\item Compute the decomposition of $\mathbf{B}_t= \bar{\bU}_t \bLambda_t \bar{\bU}^{\mathsf T}_t  $, form the product $\hat{\bU}_t=\bPsi_t \bar{\bU}_t$, and publish $ \widehat{\bU}_t \bLambda_t \widehat{\bU}^{\mathsf T}_t  $.
\end{enumerate}
\end{description}
}
\end{minipage}
}
\caption{The Mechanism for $k$-rank Approximation} \label{fig:lsi} 
\end{figure}
\begin{theorem} \label{thm:lsi}  \label{thm:lsi_spectral}
	Let $\sigma_1\geq \cdots \geq \sigma_{{\sf rk}(\bA)}$ be the singular values of $\bA$. Then for an over-sampling parameter $p$ with the most common choice being $p=k+1$, there is a single-pass mechanism that computes $k$-rank approximation $\bar{\bA}$ using  $O(k(n+d) \alpha^{-1} \kappa)$ bits while preserving $(\varepsilon, \delta)$-differential privacy such that	
	\begin{align*} 
	(i) \quad \| \bA - \bar{\bA} \|_F  & \leq 
	 \paren{1 + \frac{k}{p-1}}^{1/2} \min_{rk(A')<k}\|\bA-\bA' \|_F + \frac{ 2k}{\varepsilon}\sqrt{\frac{(n+d) \ln (k/\delta) }{p}} , \qquad \text{and} \\
	(ii) \quad \| \bA - \bar{\bA} \|_2  & \leq 
	\paren{1 + \frac{k}{p-1}}^{1/2} \sigma_{k+1}  + \frac{e\sqrt{(k+p) \sum_{j>k} \sigma_j^2}}{p} + \frac{ 2\sqrt{k(n+d) \ln (k/\delta) }}{\varepsilon} .
	\end{align*} 
\end{theorem}
The most common choice of $p$ is $k+1$, which is what we use in Table~\ref{table}.

\begin{full}
\begin{proof} 
The space complexity is easy to follow from our convention of bit complexity and because we need to store the matrix $\bOmega$ and the sketch. The privacy guarantee follows from Theorem~\ref{thm:PSG} and noting that all the singular-values of the matrix on which $\bOmega$ is operated from the right is greater than the threshold required for the statement of the Theorem~\ref{thm:PSG}, and the distribution of $\bPsi^{\mathsf T} \bOmega$ is the same as that of the second variant as we reuse $\bOmega$ (this follows from~\cite{BP08}). Now, it follows from the proof of the second variant (\thmref{PSG}) that it does not matter if we multiply $\bA$ (or $\bA^{\mathsf T}$) from left (or right, respectively) of vectors $\mathbf{g}_1, \cdots, \mathbf{g}_r$, i.e., $\sum_{i=1}^r \bA \ket{\mathbf{g}_i} \bra{\mathbf{g}_i}=\bA\bPhi$ and  $\sum_{i=1}^r \ket{\mathbf{g}_i} \bra{\mathbf{g}_i}\bA^{\mathsf T} = \bPhi \bA^{\mathsf T}$ have the same distribution. Combining all these arguments, we have the distribution of the second step of projection stage is identical to the second variant in~\figref{psg}, modulo some deterministic computation. Since, any arbitrary post-processing preserves differential privacy, we can now complete the proof by invoking~\thmref{PSG}. The privacy guarantee due to~\thmref{PSG} requires the minimum singular value to be greater than ${ \frac{ 4\sqrt{k \log (2/\delta)} \log(k/ \delta)}{\varepsilon}}$ for the first variant and ${ \frac{k \log(k/ \delta)}{\varepsilon}}$ for the second variant. By our choice of $w$, the singular values of the streamed matrix to the algorithm for $\mathsf{PSG}$ are at least the eigenvalues of $\sqrt{w^2 \I + \bA^{\mathsf T} A}$, which are all greater than $\frac{16 k \ln(k/\delta)}{\varepsilon}$. Since ${ \frac{ 4\sqrt{k \log (2/\delta)} \log(k/ \delta)}{\varepsilon}} \ll {16 \frac{k \log(k/ \delta)}{\varepsilon}}$, the privacy guarantee follows from~\thmref{PSG}.

In more detail, it follows from Bura and Pfeiffer~\cite{BP08}\footnote{We use the following result of Bura and Pfeiffer~\cite{BP08}:  for a random $n \times k$ normal matrix $\bOmega$, for large enough $n$, the vector formed by the entries of its left singular matrix is distributed normally with covariance matrix $(\bD^{-1}\bR \otimes \I)  (\bR \bD^{-1} \otimes \I)^{\mathsf T}$, where $\bOmega = \mathbf{L} \bD \bR^{\mathsf T}$.} and the proof in~\thmref{PSG} that, for $\bPhi= \sum_{i=1}^r \ket{\mathbf{g}_i} \bra{\mathbf{g}_i}$ and  large enough $n$, the distribution of $\bPsi^{\mathsf T} \bOmega$ is 
\[ \frac{ \exp (-\tr (\bA^{-1}(\bD^{-1}\bR)(\bD^{-1}\bR)^{\mathsf T} \bPhi )/2) \Delta( \bPhi (\bD^{-1}\bR)(\bD^{-1}\bR)^{\mathsf T})^{(n-r-1)/2}} {2^{rn/2} \pi^{n(n-1)/4} \Delta(\bA)^{r/2} \prod_{i=1}^r \Gamma((n-i+1)/2)}. \]

The proof for the first expression of~\eqnref{main} is as before. Following the rest of the steps of~\thmref{PSG}, using~\lemref{hermitian}, we compute for fixed $j \in [k+p]$, instead of equation~(\ref{eqn:final}), the second expression in~\eqnref{main} is bounded by the following expression
\begin{equation*}
 \left| \tr   \paren{ \bV \bSigma^{-1} \bU^{\mathsf{T}} {\mathbf{e}_i} { \mathbf{v}^{\mathsf T}}  \tilde{\bV} \bLambda^{-2} \tilde{\bV}^{\mathsf{T}} {\mathbf{\alpha}_j}  {\mathbf{\alpha}_j^{\mathsf T}}  } \tr \paren{(\bD^{-1}\bR)(\bD^{-1}\bR)^{\mathsf T}} \right| .\label{eqn:finalB}
\end{equation*}

Since $\bR$ is an orthonormal matrix, using the fact that $\tr((\bD^{\mathsf T}\bD)^{-1}) = \tr((\bOmega^{\mathsf T}\bOmega)^{-1})$, Lemma~\ref{lemma:wishart} with $p=k+1$, and following the remaining steps of~\thmref{PSG}, we have the privacy result for computing $\mathbf{B}$. The complete proof follows from the discussion stated at the start of this proof.

\begin{remark} We take the liberty to diverge a little to understand the intuition behind Bura and Pfeiffer~\cite{BP08}. One may skip this part without effecting the readability of the rest of this section. The result of Bura and Pfeiffer~\cite{BP08}  uses advance statistical tools, but the intuition can be argued using some basic statistics. On the other hand, it is well known that the singular values of random matrices are notoriously hard to compute (see~\cite{RV09}).  The basic reasoning behind their proof is the following line of argument. Since the entries of an $n \times r$ matrix $\bOmega$ is $\cN(0,1)$, then for any orthogonal matrices $\mathbf{G} \in \R^{n \times n}$ and $\bR \in \R^{r \times r}$, the entries of $\mathbf{G} \bOmega \bR^{\mathsf T}$ is also i.d.d. normal. This can be seen by translating to the vector form of the matrix, i.e., $\mathbf{v}=\mathsf{vec}(\bV)$ represent $rn$ vector with entries $\mathbf{v}_{i+nj}=\bV_{ij}$. Then $\mathsf{vec}(\mathbf{G} \bOmega \bR^{\mathsf T}) = (\mathbf{G} \otimes \bR) \mathsf{vec}(\bOmega)$. Now $\mathbf{G} \otimes \bR$ is also an orthogonal matrix, and multivariate Gaussian distribution is preserved if one multiply by an orthogonal matrix. Therefore, the distribution of the left singular vectors of $\bOmega$ is  the same as $\mathbf{G}  \bOmega  \bR^{\mathsf T}$. Consequently, for large enough $n$, the distribution of each singular vector is also spherically distributed.  
\end{remark}

\paragraph{\scshape Utility Guarantee.} In~\secref{lsi}, we showed that the mechanism for symmetric matrices does what~\cite[Section 1.2]{HMT11} prototype algorithm achieves. To construct a mechanism for  non-symmetric matrices, we  construct two sketches $\mathbf{Y}_t$ and $\bar{\mathbf{Y}}_t$ corresponding to $\bA_1$ and $\bA_2$ using a single pass over $\bA$ and using two Gaussian matrices $\bOmega$ and $\bar{\bOmega}$ of appropriate dimension, where $\bA_1=\begin{pmatrix} w\I & A \end{pmatrix}$ and $\bA_2=\begin{pmatrix} \bA^{\mathsf T} &  w\I \end{pmatrix}$ for appropriate dimension identity matrices in both  $\bA_1$ and $\bA_2$. Basically, we do the following using a single pass over the matrix $\bA$:
\begin{align*} \mathbf{Y} := \begin{pmatrix}  \mathbf{Y}_1 \\ {\mathbf{Y}}_2 \end{pmatrix} =	\begin{pmatrix} w \I_n & \bA \\ \bA^{\mathsf T} & w \I_d  \end{pmatrix} \begin{pmatrix}  \bOmega_1 \\ {\bOmega}_2 \end{pmatrix},  \end{align*} where $\I_n$ is an $n \times n$ identity matrix. 

Since $\begin{pmatrix}  \mathbf{Y}_1 & {\mathbf{Y}}_2 \end{pmatrix}^{\mathsf T}$ corresponds to a symmetric matrix, we can use the steps used in the projection stage in~\figref{lsi}. Note that Clarkson and Woodruff~\cite{CW09} compute $\bA^{\mathsf T}A$ in a single-pass over the matrix $\bA$.


Keeping the most general case in mind, we first show that the left singular vectors have hardly any role to play in bounding the perturbation. We assume that we perform $\mathsf{SVD}$. Let the $\mathsf{SVD}$  of $\bA$ be $\bU \bSigma \bV^{\mathsf T}$. In the following discussion, we compute the approximation of $\begin{pmatrix} w \I & \bA \end{pmatrix}$ and denote it by $\bA$. This is because $\begin{pmatrix} w\I & \bA \end{pmatrix}$ is more manageable and any upper bound on the approximation of this matrix is an upper bound on the approximation of the original matrix. The actual bound as computed in~\figref{lsi}  can be obtained by simply performing the computation on the singular value decomposition as performed in the last step of mechanism and using the sub-additivity of norms. 
From the discussion above, we know that $\bar{\bA}=\Pi_{\mathbf{Y}}\bA$; therefore, we need to bound $\| (\I - \Pi_{\mathbf{Y}})\bA\|$, where, unless specified, in this section $\| \cdot \|$ refers to both the Frobenius as well as the spectral norm. From the H\"{o}lder's inequality on the second moment, we have
\begin{equation} 
 \E [\| (\I-\Pi_{\mathbf{Y}}) A\|_F] \leq \paren{\E \sparen{\| (\I-\Pi_{\mathbf{Y}})A \|^2_F}}^{1/2}. \label{eq:bound1}
\end{equation} 

We now bound $\parallel (\I-\Pi_{\mathbf{Y}})A \parallel$. Let $\bLambda = \sqrt{\bSigma^2+ w^2 \I} $.  Let  $\bLambda_1$ denote the diagonal matrix formed by the first 
$k$ singular values and $\bLambda_2$ be the diagonal matrix for the rest of the singular values. We decompose $\bV^{\mathsf T}  $ similarly. Let the matrix formed by the 
first $k$ rows of $\bV^{\mathsf T}  $ be $\bV_1^{\mathsf T}  $ and by the rest of the rows be $\bV_2^{\mathsf T}  $. 

\paragraph{Left singular vectors have essentially no role in the approximation bound.} Recall that $\bOmega$ is an $2n \times k$ matrix; therefore, ${\mathbf{Y}}=\bA \bOmega= \bU \begin{pmatrix} \bLambda_1 \bV_1^{\mathsf T}   \bOmega & \bLambda_2 \bV_2^{\mathsf T}   \bOmega \\  \end{pmatrix}^{\mathsf T}  .$
It would be useful to consider the first $k$ rows of $\mathbf{Y}$ to be the one that mimics the action of $\bA$ and the rest of the rows of $\mathbf{Y}$ as a small perturbation that we wish to bound. 
We first prove that the left singular vectors have essentially no role to play in bounding the error. Let $\bA'=\bU \bA$, then the following chain of equalities are straightforward.
\begin{align} \| (\I - \Pi_{\mathbf{Y}}) \bA \| = \| \bU^{\mathsf T}  (\I - \Pi_{\mathbf{Y}}) \bA \| = \| \bU^{\mathsf T}  (\I - \Pi_{\mathbf{Y}}) \bU \bA' \| = \| \I \bA' - \bU^{\mathsf T}   \Pi_{\mathbf{Y}} \bU \bA' \|.  \label{eq:left}\end{align}

Now note that the projection matrix corresponding to a matrix $\mathbf{Y}$ is uniquely defined by $\range(\mathbf{Y})$, the range of $\mathbf{Y}$. Therefore, 
$  \range(\bU^{\mathsf T}  \Pi_{\mathbf{Y}} \bU) = \bU^{\mathsf T}   \range (\Pi_{\mathbf{Y}}) = \range (\bU^{\mathsf T}\Pi_{\mathbf{Y}}).$

Therefore, $\|  \bA' - \bU^{\mathsf T}   \Pi_{\mathbf{Y}} \bU \bA' \| = \| (\I - \Pi_{\bU^{\mathsf T}   \mathbf{Y}})\bA' \|.$ A useful way to understand the above expression is to view this geometrically and recall that unitary are just rotation in the space: projection by an unitary, followed by any projection operator, followed by the inverse of unitary brings us to the same space as projection by an operator followed by the inverse of the unitary.

\paragraph{Finding and bounding an appropriate perturbed matrix.}  We now use the identity that, for two operators $\mathbf{O}_1$ and $\mathbf{O}_2$, if the range of $\mathbf{O}_1$ is a subset of the range of $\mathbf{O}_2$, then the projection of any matrix using $\mathbf{O}_1$ will have all its norm smaller than the projection by $\mathbf{O}_2$. More concretely, we find a matrix $\mathbf{C}$ such that its range is a strict subset of the range of $\bU^{\mathsf T}  \mathbf{Y}$. We obtain this matrix by flattening out the first $k$ rows of $\bU^{\mathsf T}  \mathbf{Y}$. This is in correspondence with our earlier observation that the first $k$ rows mimic the action of $\bA$ and other rows are the perturbation that we wish to bound. Since the first $k$ rows of $\bU^{\mathsf T}  \mathbf{Y}$ is $\bLambda_1 \bV_1^{\mathsf T}   \bOmega$, let $\mathbf{C}:= \bU^{\mathsf T}   \mathbf{Y} \bOmega^{-1} \bV_1\bLambda_1^{-1} .$ The rows corresponding to the perturbation are $\bLambda_2 \bV_2 \bOmega $. Thus,  $\mathbf{C} = \begin{pmatrix} \I & \bLambda_2 \bV_2^{\mathsf T}   \bV_1 \bLambda_1^{-1} \end{pmatrix}^{\mathsf T}  .$ 

Let us denote by $\mathbf{S}= \bLambda_2 \bV_2^{\mathsf T}   \bV_1 \bLambda_1^{-1} .$ It is not difficult to see that $\range(\mathbf{C}) \subset \range(\bU^{\mathsf T}  \mathbf{Y})$. Moreover, $\Pi_{\mathbf{C}} \preceq \I$, $\Pi_{\bU^{\mathsf T}  \mathbf{Y}} \Pi_{\mathbf{C}} \Pi_{\bU^{\mathsf T}  \mathbf{Y}} \preceq \Pi_{\bU^{\mathsf T}  \mathbf{Y}}.$ This follows from the fact that $\range(\mathbf{C}) \subset \range(\bU^{\mathsf T}  \mathbf{Y})$ and the following derivation
\begin{align*} \Pi_{\bU^{\mathsf T}  \mathbf{Y}} \succeq \Pi_{\bU^{\mathsf T}  \mathbf{Y}} \Pi_{\mathbf{C}} \Pi_{\bU^{\mathsf T}  \mathbf{Y}} = \Pi_{\mathbf{C}} \Pi_{\bU^{\mathsf T}  \mathbf{Y}} = (\Pi_{\bU^{\mathsf T}  \mathbf{Y}} \Pi_{\mathbf{C}})^{\mathsf T}   = \Pi_{\mathbf{C}}. \end{align*}

An immediate result of the above is the following:
\begin{align}  \| (\I  - \Pi_{\bU^{\mathsf T}  \mathbf{Y}}) \bA' \| \leq \| (\I  - \Pi_{\mathbf{C}}) \bA' \|. \label{eq:perturb1} \end{align}

Since, $\Pi_{\mathbf{C}} =\mathbf{C} (\mathbf{C}^{\mathsf T}  \mathbf{C})^{-1} \mathbf{C}^{\mathsf T} $, we have the following set of derivations.
\begin{align*}
	\Pi_{\mathbf{C}} & = \begin{pmatrix} \I \\  \mathbf{S} \end{pmatrix} 
		\sparen{ \begin{pmatrix} \I & \mathbf{S}^{\mathsf T}   \end{pmatrix} \begin{pmatrix} \I \\ \mathbf{S} \end{pmatrix} }^{-1}
		\begin{pmatrix} \I & \mathbf{S}^{\mathsf T}   \end{pmatrix} 
		= \begin{pmatrix} \I \\  \mathbf{S} \end{pmatrix} 
		\sparen{ \begin{pmatrix} \I + \mathbf{S}^{\mathsf T}  \mathbf{S} \end{pmatrix} }^{-1}
		\begin{pmatrix} \I & \mathbf{S}^{\mathsf T}   \end{pmatrix} \\
		&= \begin{pmatrix} \I (\I +\mathbf{S}^{\mathsf T}  \mathbf{S})^{-1} \\  \mathbf{S}(\I +\mathbf{S}^{\mathsf T}  \mathbf{S})^{-1} \end{pmatrix} 
		\begin{pmatrix} \I & \mathbf{S}^{\mathsf T}   \end{pmatrix} 
		= \begin{pmatrix} (\I +\mathbf{S}^{\mathsf T}  \mathbf{S})^{-1}  & (\I +\mathbf{S}^{\mathsf T}  \mathbf{S})^{-1} \mathbf{S}^{\mathsf T}   \\  \mathbf{S}(\I +\mathbf{S}^{\mathsf T}  \mathbf{S})^{-1}  & \mathbf{S}(\I +\mathbf{S}^{\mathsf T}  \mathbf{S})^{-1} \mathbf{S}^{\mathsf T}   \end{pmatrix} \\
		& \succeq \begin{pmatrix} (\I - \mathbf{S}^{\mathsf T}  \mathbf{S})  & (\I +\mathbf{S}^{\mathsf T}  \mathbf{S})^{-1} \mathbf{S}^{\mathsf T}   \\  \mathbf{S}(\I +\mathbf{S}^{\mathsf T}  \mathbf{S})^{-1}  & 0 \end{pmatrix}, 
\end{align*}
where the last inequality uses the fact that $\I-\mathbf{S}^{\mathsf T}  \mathbf{S} \preceq (\I +\mathbf{S}^{\mathsf T}  \mathbf{S})^{-1}$ and $\mathbf{S}(\I +\mathbf{S}^{\mathsf T}  \mathbf{S})^{-1} \mathbf{S}^{\mathsf T}   \succeq 0$. Therefore, 
\begin{align*} \I - \Pi_{\mathbf{C}} \preceq \begin{pmatrix} \mathbf{S}^{\mathsf T} \mathbf{S} &  \I - (\I +\mathbf{S}^{\mathsf T}  \mathbf{S})^{-1} \mathbf{S}^{\mathsf T}  \\
							\I - ((\I +\mathbf{S}^{\mathsf T}  \mathbf{S})^{-1} \mathbf{S}^{\mathsf T})^{\mathsf T} & \I \end{pmatrix}. \end{align*}

Conjugating $\I - \Pi_{\mathbf{C}}$ with $\bLambda$, and applying the  fact that for every positive definite matrix, $P= \begin{pmatrix} \mathbf{X} & \mathbf{Y} \\ \mathbf{Y}^{\mathsf T} & \mathbf{Z} \end{pmatrix},$ we have $\| P \| \leq \|\mathbf{X}\| + \|\mathbf{Z}\|$, we get 
\begin{align}  \| (\I  - \Pi_{\mathbf{C}}) \bA' \| \leq \|\mathbf{S}^{\mathsf T}  \mathbf{S} \bA'\| + \| \bA' \| \label{eq:perturb} \end{align}
for any norm.
From here on, it is easy arithmetic to show that 
\begin{equation} \| (\I  - \Pi_{\mathbf{C}}) \bA' \| \leq \sqrt{\| \bLambda_2' \| + \| \bLambda_2' \bV_2^{\mathsf T}   \bOmega (\bV_1^{\mathsf T}   \bOmega)^{-1} \|} \label{eq:bound2}
\end{equation} 
for both the required norms. Using~\eqnref{left},~\eqnref{perturb1} and~\eqnref{perturb}, this gives us a bound on the approximation of matrix $\bA'$.
Till this point, our analysis closely follows the ideas of~\cite{HMT11}, accommodating the steps of our algorithm. Now, all that remains is to bound $\| \bLambda_2 \bV_2^{\mathsf T}   \bOmega (\bV_1^{\mathsf T}   \bOmega)^{-1} \|$, and for this, we have to analyze the  matrix $\bOmega$. 

\paragraph{\scshape Error bound for Frobenius norm}
We  now exploit the rotational invariance of a Gaussian distribution. An important point to note is that  $(\bOmega \bOmega^{\mathsf T})^{-1}$ exists and  has a well defined trace. The first part of the right hand side of~\eqnref{bound2} is immediate. Thus, if we bound $\E [\| \bLambda_2' \V_2^{\mathsf T}   \bOmega (\bV_1 \bOmega)^{-1} \|]$, we are done. This could be accomplished as below.
\begin{align*}
	\E [\| \bLambda_2 \V_2^{\mathsf T}   \bOmega (\bV_1 \bOmega)^{-1} \|] 
									&\leq \sqrt{\E \sparen{\sum_{ij} | (\bLambda_2)_{ij}' \Pi_{ij} (\bV_1\bOmega ^{-1})_{jj} |  }} \\
									&\leq \sqrt{\| \bLambda_2' \|_F \| \bOmega^{-1} \|_F}  \\
									&= \sqrt{\| \bLambda_2' \|_F \mathsf{Tr} \paren{ \paren{ \bOmega \bOmega^{-1}}^{\mathsf T}   } } 
									= \sqrt{\| \bLambda_2' \|_F \mathsf {Tr} (\bOmega \bOmega^{\mathsf T}  )^{-1} } \\
									& \leq \sqrt{\mathsf{Tr} (\bOmega \bOmega^{\mathsf T}  )^{-1}} \min_{rk(\bA')\leq k}\| \bA - \bA'\|_F + \sqrt{(n+d)w \mathsf{Tr} (\bOmega \bOmega^{\mathsf T}  )^{-1}}.
\end{align*}

The utility guarantee follows by plugging this value in~\eqnref{bound2}, and combining~\eqnref{bound1} and the fact that $(\bOmega \bOmega^{\mathsf T})^{-1}$ has a well defined trace $k/(p-1)$~\cite{Muirhead}.

\paragraph{\scshape Error bound for Spectral norm}
In order to bound the second term, we use few well known facts in the theory of random matrices to simplify~\eqnref{bound2}.
In particular, using Lemma~\ref{lemma:random1} and~\ref{lemma:random2}, and Holder's inequality, the statement of the theorem for the spectral norm follows. The utility bound then follows using the same arithmetic of representing $\bLambda'$ in terms of $\bLambda$ as done in the case of Frobenius norm. In more details, we first bound 
\begin{align*} 
\E \sparen{\| \bLambda_2' \bV_2^{\mathsf T}   \bOmega (\bV_1^{\mathsf T}   \bOmega)^{-1} \|} & \leq \| \bLambda_2' \|  \paren{ \E \sparen{ \| (\bV_1^{\mathsf T}   \bOmega)^{-1} \|^2_F \| } }^{1/2} + \| \bLambda_2' \|_F \paren{ \E \sparen{\| (\bV_1^{\mathsf T}   \bOmega)^{-1} \| }} \\
	& \leq \| \bLambda_2' \|  \paren{ \E \sparen{ \|  \bOmega^{-1} \|^2_F \| } }^{1/2} + \| \bLambda_2' \|_F \paren{ \E \sparen{\| \bOmega^{-1} \| }}, 
\end{align*}
 and then invoke Lemma~\ref{lemma:random2} followed by the sub-additivity of norms. Making these substitution and on simplification, we get the bound stated in~\thmref{lsi}(ii).
\[ (\ref{eq:bound2})  \leq \paren{1 + \frac{k}{p-1}}^{1/2} \| \bLambda_2 \|_2  + \frac{e\sqrt{(k+p) \sum_{j>k} \lambda_j^2}}{p} + \frac{ 2\sqrt{k(n+d) \ln (2/\delta) }}{\varepsilon}. \]


\subsubsection{Tightness of Bounds and Comparison with Earlier Works} \label{sec:comparison}
We compare our results with the best possible results in the non-private setting. Eckart and Young~\cite{EY36} have shown that the quantity $\min_{rk(\bA')<k}\|\bA-\bA' \|_F$ in the first term of~\thmref{lsi}(i) is optimal. Likewise, Mirsky~\cite{Mirsky} proved that $\lambda_{k+1}$ is the minimum spectral error for $k$-rank approximation. The second term in~\thmref{lsi}(ii) shows that we also pay for the Frobenius norm error when doing a unified analysis. However, when the oversampling parameter $p \approx k$, then the factor on $\lambda_{k+1}$ is constant and that on the second term is of order $k^{-1/2}$. In fact, on closer analysis, 
\[ \| \bA - \bar{\bA} \|_2   \leq 
	\paren{1 + \frac{k}{p-1} + \frac{e (\sqrt{(k+p) \min\{d,n \} - k})}{p}} \lambda_{k+1} +  \frac{ 2\sqrt{k(n+d) \ln (2/\delta) }}{\varepsilon}, \]
	therefore, the error always lies within some polynomial factor of $\lambda_{k+1}$, modulo some additive  error. As pointed out of Halko {\it et al.}~\cite{HMT11}, one can improve this by power-iteration, the method used by Hardt and Price~\cite{Hardt13}~and Hardt and Roth~\cite{HR13} and multiple pass mechanism. However, it seems unlikely to improve it in a single-pass. 

Kapralov and Talwar~\cite{KT13} showed a lower bound on additive error  when $\delta=0$ for neighbouring data differing by unit spectral norm. Our privacy proof depends strongly on the fact that $\delta \neq 0$. In fact, our bound is vacuous if $\delta=0$. Though incomparable due to difference in the notion of neighbouring data, this separation gap further strengthen the belief that better bounds are possible for $\delta \neq 0$. Recently, Dwork {\it et al.}~\cite{DTTZ14} also showed a bound in the online learning model, which is a different model of computation.

We compare our results with the works stated in Table~\ref{table} in more detail. 
\begin{description}
	\item [Chaudhary {\it et al.}~\cite{CSS12}:] They give low-rank approximation in the spectral norm. Additionally, they achieve $(\varepsilon, 0)$-differential privacy, which is only achieved by Kapralov and Talwar~\cite{KT13}. Their definition of neighbouring data sets can be (arguably) considered the most general in the sense that they consider two data sets neighbouring if they differ by at most one in the spectral norm. They use exponential distribution to sample a singular vector and give a heuristic, but practical implementation using Markov chain Monte-Carlo. On the negative side, their mechanism uses $k$ rounds; therefore, it cannot be implemented in a streaming fashion. Since the notion of neighbouring data-sets and privacy guarantee achieved is different from that of ours, we believe our result is incomparable to that of Chaudhary {\it et al.}~\cite{CSS12}. However, if we just concentrate on the additive error bound, they achieve a bound of order $ O(nk/\varepsilon)$ compared to our bound  $\tau \leq \frac{ 2\sqrt{k(n+d) \ln (2/\delta) }}{\varepsilon}$~(\thmref{lsi}(ii)).
	\item [Hardt and Roth~\cite{HR12}:] The authors  use two passes over the input matrix; therefore, it does not fall in our one-pass streaming model of computation. They use the same notion of neighboring data-sets as we do in this paper. This makes their coherence conditions and notion of neighbouring data sets rotationally invariant. As argued by Blocki {\it et al.}~\cite{BBDS12}, we achieve a better utility bound in the range finding step. Intuitively, this could be seen as a consequence of the absence of additive Gaussian noise. More concretely, Hardt and Roth~\cite{HR12} achieved an error bound of ${\sqrt{kn}\log(k/\delta)}/{\varepsilon} + \sqrt{{\mu \|\bA\|_F (n/d)^{1/2}\log(k/\delta)}/{\varepsilon}}$. Their error bound depends on $\| \bA \|_F$, which can be as large as $\sqrt{nd}$ for binary matrices in the worst case when the matrix is not as well-behaved as captured by low-coherence assumption. On the other hand, we achieve a bound that is independent of $\|\bA\|_F$.
	\item [Hardt and Roth~\cite{HR13}:] In some sense, this paper is based on Krylov subspace iteration combined with powering method of Halko {\it et al.}~\cite{HMT11}. They define two data-sets as neighbouring in the same manner as in Hardt and Roth~\cite{HR13}.  The coherence definition used in this paper depends on the maximum value of the left or right singular vectors, and is, therefore, rotationally variant. This work assumes that the singular value are well separated, i.e., the first and the $k$-th singular value has a non-trivial separation, and give $\lra$ in spectral norm. Their bound, however, depends on the rank of the input matrix. Their mechanism uses $k$ rounds of subspace generation, each of which depends on the spectrum of the matrix and uses the power-iteration method~\cite{HMT11}. This helps them in achieving better multiplicative bound,  but make them unsuitable in a streaming model. A note on multiplicative bound is due here. We believe that the general application of $\lra$ is for thin matrices with very small tail singular values (for example, see,~\cite{WS00}). Therefore, we feel that in practical scenario, polynomial multiplicative error would not be that big an issue. The additive error bound computed by~\cite{HR13} $\tau \leq O(k^2 \varepsilon^{-1} \sqrt{ (\mathsf{rank}(A) \mu + k\log n) \log (1/\delta)} \log n)$ compared to  $\tau \leq \frac{ 2\sqrt{k(n+d) \ln (2/\delta) } }{\varepsilon}$ (\thmref{lsi}(ii)).
	\item [Kapralov and Talwar~\cite{KT13}:] The only assumption this paper makes is that of singular value separation of the same form as in Hardt and Roth~\cite{HR13}. They also give low-rank approximation in the spectral norm. Additionally, they achieve $(\varepsilon, 0)$-differential privacy, which is only achieved by Chaudhary {\it et al.}~\cite{CSS12}. They use the same definition as in Chaudhary {\it et al.}~\cite{CSS12}. They also sample a singular vector from exponential distribution, but they give a net-based algorithm to perform the sampling in polynomial time. On the negative side, their mechanism uses $k$ rounds; therefore, it cannot be implemented in a streaming fashion. Since the notion of neighbouring data-sets and privacy guarantee achieved is different from that of ours, we believe our result is incomparable to that of~\cite{KT13}. However, if we just concentrate on the additive error bound, they achieve a bound of $ O(dk^3/(\varepsilon \gamma^2))$, where $\gamma$ is the separation between the singular values, compared to our bound  $\tau \leq \frac{ 2\sqrt{k(n+d) \ln (2/\delta) }}{\varepsilon}$~(\thmref{lsi}(ii)).
	\item [Hardt and Price~\cite{Hardt13}:] In this recent work, Hardt and Price~\cite{Hardt13} gave a robust subspace iteration mechanism that allows to publish $\lra$ with noise independent of the rank of the input matrix, thereby, resolving one of the open problems in~\cite{HR13}. They define neighbouring  data-sets   in the same manner as in~\cite{HR12,HR13}. However, they also make an assumption on the singular value separation--a separation between the $k$-th and $(k+1)$-th singular value of the input matrix. Their mechanism uses $k$ rounds of subspace generation, each of which depends on the spectrum of the matrix to reduce the multiplicative error; therefore, it cannot be implemented in a streaming fashion. We achieve a bound of  $\frac{ 2\sqrt{k(n+d) \ln (2/\delta) }}{\varepsilon}$~(\thmref{lsi}(ii)) compared to ${\lambda_1 \sqrt{kn \mu \log(1/\delta) \log(n/\gamma)\log \log (n/\gamma)}}/{\varepsilon \gamma^{1.5} \lambda_k}$ of~\cite{Hardt13}.
	\item [Dwork {\it et al.}~\cite{DTTZ14}:] Dwork {\it et al.}~\cite{DTTZ14}  gave the first single-pass online learning algorithm for private low-rank approximation under the assumption that the rows of the input matrix are normalized. They consider the online-learning model, which is very different from our model, and we do not see any natural way to compare. They use the {\it follow the perturbed leader} ($\mathsf{FTL}$) algorithm of Kannan and Vempala~\cite{KV05} with the binary tree technique of Dwork {\it et al.}~\cite{DNPR10}. This idea was previously used by Jain {\it et al.}~\cite{JKT12} as well. They give a bound that assumes a lower bound of $k \sqrt{n} \log^2(m/\delta)/\varepsilon^2$ on the optimal value, where $\delta < 1/ \poly(n)$. More concretely, if $\mathsf{OPT}$ is the optimal value, then their error bound is $O(\sqrt{k \mathsf{OPT}} n^{1/4} \log^2(m/\delta))$. We do not make any of the assumptions made by them and, if we just consider the end result, we achieve a bound which is factor $k\sqrt{n}$ better than theirs (see~\thmref{lsi}(ii)). The case that we are able to bypass their lower bound  gives a mathematical indication that the unit norm notion of neighbouring data is strictly weaker than user-level privacy.
\end{description}

\end{proof}
\end{full}

\begin{extended}
\noindent {\em Proof Sketch.} We need the privacy guarantee for both $\mathsf{PSG}_1$ and $\mathsf{PSG}_2$ algorithm for our privacy proof -- $\mathsf{PSG}_1$ for the range finding stage  and $\mathsf{PSG}_2$ because we reuse $\bOmega$ in the second step of the projection stage. The range finding pushes the singular values above the threshold of~\thmref{PSG} (ii).  Since the rest of the computations are deterministic, privacy follows from Lemma~\ref{lem:post}, Theorems~\ref{thm:PSG}  and~\ref{thm:DRV},  and our choice of $w$. The space guarantee is also straightforward. Now, $\Pi_\bY$ has a decomposition $\bPsi \bPsi^{\mathsf T}$ for some orthonormal basis $\bPsi$, which also forms an orthonormal basis for the approximated matrix $\bar{A}$ after the run of the algorithm. Therefore,  $\mathbf{B}$ must satisfy $\mathbf{B}\bPsi^{\mathsf T} \bOmega \approx \bPsi ^{\mathsf T} \bY$.  This is what step 2 does. Therefore, $\bar{\bA}_t =  \Pi_\bY \bA_t$ for the projection operator $\Pi_\bY$ with the same range as $\bPsi$. The rest of the utility proof  relies heavily on the fact that  $\tr (\bOmega \bOmega^{\mathsf T})^{-1}$ is bounded,  left singular vectors of $\bA$ does not play any essential role in the concentration bound, and the rotational invariance of Gaussian vectors.  We  bound the norm of $\| (\I-\Pi_\bY) \bA\|$ for both the spectral and Frobenius norm.  Once we have a bound on  $\| (\I-\Pi_\bY) \bA \|$, we use standard results in random matrix theory to get the final bounds. 
This approach is completely different from that of Sarlos~\cite{Sarlos06} and Clarkson and Woodruff~\cite{CW09}, where the authors use the bound computed in the estimate of \mult. Here we use perturbation theory, more aligned with~\cite{DTTZ14}~and~\cite{HMT11}.  The detail proof appears in~\appref{lra}.

We perform a detailed comparison of our result with earlier works  in Appendix~\ref{sec:comparison} though we make note on $\alpha$ and $\tau$ here. 
Our additive error $\tau$ almost meets the lower bounds of Hardt and Roth~\cite{HR13} for no coherence assumption.  For $\alpha$, Eckart and Young~\cite{EY36} have shown that the first term of~\thmref{lsi}(i) is optimal, and Mirsky~\cite{Mirsky} proved that $\sigma_{k+1}$ is the minimum spectral error for $k$-rank approximation. It can be easily shown that $\alpha$ in~\thmref{lsi}(ii) is polynomial factor of $\sigma_{k+1}$. One can further improve it by using power-iteration as done by Hardt and Price~\cite{Hardt13}~and Hardt and Roth~\cite{HR13} at the expense of multiple passes over the private matrix $\bA$; however, it seems unlikely that one can improve it in one pass. 
\end{extended}

\section{Other Applications of  Private Sketch Generation} \label{sec:application}
In this section, we give two applications of $\mathsf{PSG}_1$: mechanisms for \mult \ and \linear.~\thmref{PSG} guarantees privacy if the singular values are high enough, so our basic approach would be to lift the singular values of the private matrix above the threshold of~\thmref{PSG}(i). However, we have to be careful. For example, if we use the affine transformation based approach of Blocki {\it et al.}~\cite{BBDS12} and Upadhyay~\cite{Upadhyay13}, then it would  lead to  an additive error  proportional to the Frobenius norm of input matrices. To control the additive error, we follow a different approach. We first transform any conforming matrices $\bA$ and $\mathbf{B}$ to  $\bA'$ and $\mathbf{B}'$, respectively, and then use the identity, $\begin{pmatrix} \I & \bA'  \end{pmatrix}\begin{pmatrix} \I & \mathbf{B}'  \end{pmatrix}^{\mathsf T} = \begin{pmatrix} \I+\bA'\mathbf{B}'^T  \end{pmatrix}$ to perturb the input matrix with a careful choice of parameters. Intuitively, $\tau$ is due to the identity term of the published matrix. We use the same idea for \linear \ as well.  



\subsection{Matrix Multiplication} \label{sec:prod}
We present the mechanism for \mult \ below and the result is stated in~\thmref{prod}. The main idea is to lift the spectra of the input matrices above the threshold of ~\thmref{PSG}(i). 
\small{
\begin{description} 
	\item [{\sc Initialization.}] On input parameters $\alpha, \beta, \varepsilon, \delta$, set $ r=O(\log(1/\beta)/\alpha^2)$. Set $s = {\sqrt{16r \ln(\frac{2}{\delta})}}/\varepsilon \ln(\frac{16r}{\delta})$. Set the intial sketches of $\bA$ and $\mathbf{B}$ to be all zero matrices $\bY_{\bA_0}$ and $\bY_{\mathbf{B}_0}$.
	\item [{\sc Data-structure update.}]  Set $d=\max \{d_1,d_2\}$. On input a column $a$ of an $n \times d_1$ matrix $\bA$ and column $b$ of an $n \times {d_2}$ matrix $\mathbf{B}$ at time epoch $t$, set the column vector $\widehat{\bA}_{:a} = \begin{pmatrix} {s \mathbf{e}_a} & \mathbf{0}^{n+d} &   \bA_{:a} \end{pmatrix} $ and $\widehat{\mathbf{B}}_{:b} = \begin{pmatrix} s \mathbf{e}_b &   \mathbf{0}^{n+d}  &  \mathbf{B}_{:b}  \end{pmatrix} $. 
	 Invoke $\mathsf{PSG}_1$ with inputs $(\widehat{\bA}_{:a}, r, n+d)$ and $(\widehat{\mathbf{B}}_{:b}, r, n+d)$. Update the sketches by replacing the columns $a$ of $\bY_{\bA_{t-1}}$ and $b$ of $\bY_{\mathbf{B}_{t-1}}$ by the respective returned sketches to get the sketch $\bY_{\bA_t}, \bY_{\mathbf{B}_t}$.
	 \item [{\sc Answering matrix product.}] On request to compute the product at time $t$, compute $\bY_{\bA_t}^{\mathsf T} \bY_{\mathbf{B}_t}$.
\end{description}}

\begin{theorem} \label{thm:prod}
	Let $\bOmega$ be the  random matrix used by $\mathsf{PSG}$. Then, the data-structure generated by mechanism above uses  $O(d \alpha^{-2}\kappa \log(1/\beta) )$ bits of space, and on input conforming matrices $\bA$ and $\mathbf{B}$,  computes $(\alpha,\beta,\tau)$-\mult \   with  $\tau \leq s^2\sqrt{n} \alpha$ additive error and $(\varepsilon, \delta)$-differential privacy. 
\end{theorem}
\begin{full}
\begin{proof} 
The proof of the utility of mechanism in~\secref{prod} follows readily from Lemma~\ref{lemma:prod}, which is the variance bound computed by~\cite{KN14}. 
\begin{lemma} \label{lemma:prod} 
Let $\bOmega$ be a $r \times n$  matrix as constructed in~\secref{main} with every entries picked from the distribution $\cN(0,1)$, then for a set of $m$ vectors, $\mathbf{v}_1, \cdots, \mathbf{v}_m \in \R^n$, with probability at least $1-2\exp(-r\alpha^2 /8)$, for any pair $\mathbf{v}_i,\mathbf{v}_j$, we have $|\brak{\bOmega \mathbf{v}_i,\bOmega \mathbf{v}_j} - \brak{\mathbf{v}_i,\mathbf{v}_j}| \leq \alpha \| \mathbf{v}_i \| \cdot \| \mathbf{v}_j\|.$
\end{lemma}

The basic intuition is that the multiplication of the scaled identity matrix causes the additive error while the multiplicative error is due to the result of~\cite{KN14}. We follow up with the details.
 We need to upper bound the quantity $\| \bA^{\mathsf T} \mathbf{B} - \widehat{\bA}^{\mathsf T} \bOmega^{\mathsf T} \bOmega \widehat{\mathbf{B}}/s^2\|_F$. First note that  
\[  \widehat{\bA}^{\mathsf T} \bOmega^{\mathsf T} \bOmega \widehat{\mathbf{B}} =   \begin{pmatrix} s\I & 0 &0 & \bA^{\mathsf T} \end{pmatrix} \bOmega^{\mathsf T} \bOmega \begin{pmatrix}s \I & 0 &0 & \mathbf{B}^{\mathsf T} \end{pmatrix}^{\mathsf T} =  (s^2\I \bOmega^{\mathsf T} \bOmega \I + \bA^{\mathsf T} \bOmega^{\mathsf T} \bOmega \mathbf{B}).\]

Therefore, 
\begin{align}
 \| \bA^{\mathsf T}\mathbf{B} - \widehat{\bA}^{\mathsf T} \bOmega^{\mathsf T} \bOmega \widehat{\mathbf{B}}\|_F &= \| \bA^{\mathsf T}\mathbf{B} - \bA^{\mathsf T} \bOmega^{\mathsf T} \bOmega \mathbf{B} - s^2\I \bOmega^{\mathsf T} \bOmega \I \|_F \nonumber \\
 	 &\leq \| \bA^{\mathsf T}\mathbf{B} - \bA^{\mathsf T} \bOmega^{\mathsf T} \bOmega \mathbf{B} + s^2\I - s^2\I \bOmega^{\mathsf T} \bOmega \I \|_F \nonumber \\ 
	 & \leq\| \bA^{\mathsf T}\mathbf{B} - \bA^{\mathsf T} \bOmega^{\mathsf T} \bOmega \mathbf{B} \|_F + s^2\| \I - \bOmega^{\mathsf T} \bOmega \|_F. \label{eq:total}
 \end{align}

To bound the first term, let random variable $X_{ij}$ denote $({\bA^{\mathsf T}} {\mathbf{B}})_{ij} - ( {\bA^{\mathsf T}} \bOmega^{\mathsf T}  \bOmega {\mathbf{B}})_{ij}$. Then, with probability at least $1-2\exp(-k\alpha^2 /8)$, we have $|X_{ij}| \leq \alpha \| {\bA}_{:i}\|_2 \cdot \| {\mathbf{B}}_{:j}\|_2$. Using Lemma~\ref{lemma:prod}, this results in

 \begin{equation} 
 {\| {\bA^{\mathsf T}} {\mathbf{B}} - {\bA^{\mathsf T}} \bOmega^{\mathsf T} \bOmega {\mathbf{B}} \|_F}^2 = 
 		 { \sum |X_{ij}|^2} \leq 
		 {\sum \alpha^2 \| {\bA}_{i:}\|^2_2 \| {\mathbf{B}}_{:j}\|^2_2}  \leq  
		\alpha^2  \|\bA\|_F^2 \|\mathbf{B}\|_F^2. \label{eq:first}
\end{equation}



For the second term, we need to bound the variance on unitaries. This follows from the following set of inequalities.
\begin{align}
	\| \bU_1^{\mathsf T} \bOmega^{\mathsf T} \bOmega \bU_2 - \bU_1\bU_2 \|_2 &= \| \bU_1 (\bOmega^{\mathsf T} \bOmega - \I) \bU_2 \|_2 = \| \bOmega^{\mathsf T} \bOmega  - \I \|_2 \nonumber \\
							& =  \paren{ \max \frac{x^{\mathsf T} (\bOmega^{\mathsf T} \bOmega - \I)x}{\brak{x,x}} }   
							 \leq {( (1+\alpha) -1)} = {\alpha} , \label{eq:prod2}
\end{align}
where the inequality follows from Theorem~\ref{thm:JL} and noting that $r$ still satisfies the theorem requirement. 
 The result follows by adjusting the value of $\alpha$ after plugging this and~\eqnref{first} in~\eqnref{total}, using the fact  that $\| \mathbf{X}\|_2 \leq \|\mathbf{X}\|_F \leq \sqrt{n} \|\mathbf{X}\|_2$ for any $n \times n$ matrix $\mathbf{X}$.

\end{proof}
\end{full}
\begin{extended}
\noindent {\em Proof Sketch.} The space requirement of the data-structure is straightforward by the choice of $r$. The privacy guarantee follows because $ \widehat{\bA}_t  \widehat{\bA}_t^{\mathsf T}  \succeq \bU \paren{16r \ln(2/\delta)/\varepsilon} \ln(16r /\delta)^2 \I \bU^{\mathsf T} = \sigma_{\mathsf {min}}^2 \I$ at any time $t$, i.e., the singular values of the perturbed matrices are above the threshold of Theorem~\ref{thm:PSG}.  The proof of utility readily follows using the variance bound on $\| \bA^{\mathsf T} \bOmega^{\mathsf T} \bOmega \mathbf{B} - \bA \mathbf{B} \|_F^2$ by Kane and Nelson~\cite{KN14}, bounding $\| \bOmega^{\mathsf T} \bOmega - \I \|_2$ that results from taking in to account the perturbation made to the input streams in Step 1, and norm inequalities. The details are in~\appref{prod}.
\end{extended}

\noindent 
\subsection{ Linear Regression} \label{sec:linear}
Our mechanism for \linear \ is presented below and the result is stated in~\thmref{linear}. The main idea is to lift the singular values of the input matrix above the threshold of~\thmref{PSG}(i). 
\small{ 
\begin{description}
	\item [{\sc Initialization.}] On input parameters $\alpha, \beta, \varepsilon, \delta$, set $r=O(d \log(1/\beta)/\alpha)$, $s =  {\sqrt{16r  \ln(2/\delta)}} \varepsilon^{-1} \ln(16r /\delta)$, and $\bY_{\bA_0}$ to be all zero matrix. 
	\item [{\sc Data-structure update.}] On input a column $c$ of an $n \times d$ matrix  $\bA$ at time epoch $t$, set  the column vector $\widehat{\bA}_{:c} =   \begin{pmatrix}  s \mathbf{e}_c  &  \mathbf{0}^{n+d}  &  \bA_{:c}  \end{pmatrix}$. Call $\mathsf{PSG}_1$ with input $(\widehat{\bA}_{:c},  r,  n+d)$. Update the sketch of $\bA$ by replacing the column $c$ of $\bY_{\bA_{t-1}}$ by the  returned sketch to get the sketch $\bY_{\bA_t}$.
	\item [{\sc Answering queries.}] On being queried with a vector $\mathbf{b}_i$, set the column vector $\widehat{\mathbf{b}}_i=\begin{pmatrix}   \mathbf{0}^d  &  { \mathbf{0}^{n+d}}  &  \mathbf{b}_i  \end{pmatrix}$. Call $\mathsf{PSG}_1$ with input $(\widehat{\mathbf{b}}_i, r,n+d)$ to get the sketch $\bY_{\mathbf{b}_i}$. 	Compute a vector $\bx_i$ satisfying $\min_{\bx} \| \bY_{\bA_t} \bx_i-\bY_{\mathbf{b}_i}\|$.
\end{description}}
\begin{theorem} \label{thm:linear}
	Let $\bOmega$ be  $r \times 2(n+d)$ matrix used by $\mathsf{PSG}$, where $r=O(d\log(1/\beta)/\alpha)$. Then  the data-structure generated above requires $O(d^2 \alpha^{-1}\kappa \log(1/\beta))$ bits and  allows to solve $(\alpha,\beta,\tau)$-\linear \ problem in an $(\varepsilon, \delta)$-differentially private manner with   $\tau \leq O(s^2\sqrt{n} \alpha).$
\end{theorem}
\begin{full}
\begin{proof} 
The formal proof of Theorem~\ref{thm:linear} is identical to~\cite{Sarlos06}, which was refined in~\cite{CW09}, modulo the analysis to consider the lift of singular value and using the bound of~\cite{KN14} (Lemma~\ref{cor:linear}). We first start with the intuition. For this, let us recall the main result of~\cite{CW09} with matrices $\widehat{\bA}$ and $\widehat{\mathbf{B}}$. Casted with these matrices, the result of~\cite{CW09} is as follows:
\begin{theorem}
 Let $\bOmega$ be as in~\secref{linear}, $\widehat{\bA}$ and $\widehat{\mathbf{B}}$ be the matrices as constructed in the mechanism stated in~\secref{linear}. Then with probability at least $1-\beta$, $\| \widehat{\bA} \widehat{\mathbf{X}} - \widehat{\mathbf{B}} \| \leq (1+ \alpha) \| \widehat{\bA} \widetilde{\mathbf{X}} - \widehat{\mathbf{B}} \|$, where 
 $$ \widehat{\mathbf{X}} = \arg \min_{\mathbf{X}} \| \bOmega (\widehat{\bA} X - \widehat{\mathbf{B}}) \|_F^2 \qquad \widetilde{\mathbf{X}} = \arg \min_{\mathbf{X}} \|  (\widehat{\bA} X - \widehat{\mathbf{B}}) \|^2_F.   $$
\end{theorem}

Now note that $\widehat{\bA}^{\mathsf T} = \begin{pmatrix} \I & A \end{pmatrix}^{\mathsf T}$ and $\widehat{\mathbf{B}}^{\mathsf T} =  \begin{pmatrix} \I & B \end{pmatrix}^{\mathsf T}$. In other words, both $\widehat{\bA}$ and $\widehat{\mathbf{B}}$ have the same block matrices for the first few rows. Since, $\widetilde{\mathbf{X}}$ minimizes the value of $(\widehat{\bA} \mathbf{X} - \widehat{\mathbf{B}})$, it has entries $1$ except for the last $n$ positions. Therefore, one way to look at the bound is that the multiplicative error is due to the last $n$ entries of $\widehat{\bA}$ and $\widehat{\mathbf{B}}$ and the additive error is due to the approximation given by the sketch on the rest of the entries. 

We follow the approach of~\cite{CW09}, which is the refinement of~\cite{Sarlos06}. Their approach to prove the utility bound for linear regression  works in two stages and we reiterate it here. We first bound $\| \bU^{\mathsf T} \widehat{\bA} (\widetilde{\mathbf{X}} - \widehat{\mathbf{X}})\|_F$, where $\widehat{\mathbf{X}}$ is the right value of the linear regression,  $\widetilde{\mathbf{X}}$ is the value of regression for the sketch of $\widehat{\bA}$ and $\widehat{\mathbf{B}}$, and $\bU$ is left singular matrix of $\widehat{\bA}$. Let $\widetilde{\mathbf{X}}'$ represents the last $n$ entries of $\widetilde{\mathbf{X}}$ (similarly define for $\widehat{\mathbf{X}}$) and $\widetilde{\mathbf{X}}''$ be the rest of the entries (similarly for $\widehat{\mathbf{X}}$). 

The following lemma given by~\cite{KN14} is key to our analysis. This is an equivalent for~\eqnref{prod2} with this value of $r$. Note that a random Gaussian matrix of the order we considered is a Johnson-Lindenstrauss transform.
\begin{lemma} \label{cor:linear}
	Given  $r=O(d\log (1/\beta)/\alpha)$. Let $\bU$ be any unitary matrix. If $\bOmega$ satisfies the Johnson-Lindenstrauss bound, then with probability at least $1-\beta$, we have $\| \bU \bOmega^{\mathsf T} \bOmega \bU^{\mathsf T} - \I \|_2 \leq \alpha $.
\end{lemma} 

Using Lemma~\ref{cor:linear}, the bound on $\| \bU^{\mathsf T} \widehat{\bA}(\tilde{\mathbf{X}} - \widehat{\mathbf{X}})\|_F$ follows from the triangle inequality.  The remainder of the proof follows from the Pythagorus theorem and an observation that $A$ and $\bU$ have the same column-space. A structural property from our construction that we repeatedly use for moving from $\widehat{\bA}$ to $A$ (respectively, from $\widehat{\bA}$ to $B$) is that except for the block matrix formed by the last $n$ rows, the rest of the entries of the  matrices $\widehat{\bA}$ and $\widehat{\mathbf{B}}$ are identical. The details follows.

Let $B$ be the matrix formed by the set of queries $\{\mathbf{b}_1, \cdots, \mathbf{b}_m\}$ and $\bU$ be the left singular matrix of $\widehat{\bA}$. Since the columns of $\bU$ is a set of orthonormal vectors, we have $\bU \bU^{\mathsf T}\bU=\bU$ and $\| \bU^{\mathsf T} \bU \mathbf{C} \|_F=\| \bU \mathbf{C} \|_F$ for any matrix $\mathbf{C}$.
Therefore, it suffices for the utility bound to prove a bound on $\| \bU^{\mathsf T} \widehat{\bA}(\widehat{\mathbf{X}}- \widetilde{\mathbf{X}})\|$. For this, we first prove that $\bU^{\mathsf T} \bOmega^{\mathsf T} \bOmega \widehat{\bA} (\widehat{\mathbf{X}}- \widetilde{\mathbf{X}})$ has a small norm. We have
\begin{align*}
	\bU^{\mathsf T} \bOmega^{\mathsf T} \bOmega \widehat{\bA} (\widehat{\mathbf{X}}- \widetilde{\mathbf{X}}) &= \bU^{\mathsf T}\bOmega^{\mathsf T} \bOmega \widehat{\bA} (\widehat{\mathbf{X}}- \widetilde{\mathbf{X}}) + \bU^{\mathsf T} \bOmega^{\mathsf T} \bOmega (\widehat{\mathbf{B}}-\widehat{\bA} \widehat{\mathbf{X}}) \\
								&= \bU^{\mathsf T}\bOmega^{\mathsf T} \bOmega (\widehat{\mathbf{B}}-\widehat{\bA} \widetilde{\mathbf{X}}).
\end{align*}
This is because $\bU^{\mathsf T} \bOmega^{\mathsf T} \bOmega (\widehat{\bA} \widehat{\mathbf{X}} -\widehat{\mathbf{B}}) = \widehat{\bA}^{\mathsf T} \bOmega^{\mathsf T} \bOmega (\widehat{\bA}\widehat{\mathbf{X}}-\widehat{\mathbf{B}}) =0.$
Therefore, from Theorem~\ref{thm:prod} with $\alpha' = \sqrt{\alpha /d}$ (since we chose $r$ in~\thmref{prod} which differs by a factor of $\alpha$ and $1/d$ with respect to that in~\thmref{linear}), we have
\begin{align*}
	\| \bU^{\mathsf T} \bOmega^{\mathsf T} \bOmega \widehat{\bA} (\widehat{\mathbf{X}}- \widetilde{\mathbf{X}}) \|_F &= \| \bU^{\mathsf T}\bOmega^{\mathsf T} \bOmega (\widehat{\mathbf{B}}-\widehat{\bA}\widetilde{\mathbf{X}}) \|_F	\\		
									&\leq \sqrt{\alpha}  \|\mathbf{B} - \bA\widetilde{\mathbf{X}}' \|_F + \sqrt{\tau}
\end{align*}

From the sub-additivity of the norm and property of conforming matrices~(\lemref{conform}), we have 
\begin{align*}
	 \| \bU^{\mathsf T}\widehat{\bA}(\widehat{\mathbf{X}}- \widetilde{\mathbf{X}})\|_F &\leq \| \bU^{\mathsf T} \bOmega^{\mathsf T} \bOmega   \widehat{\bA} (\widehat{\mathbf{X}}- \widetilde{\mathbf{X}}) \|_F + \| \bU^{\mathsf T} \bOmega^{\mathsf T} \bOmega  \widehat{\bA}  (\widehat{\mathbf{X}}- \widetilde{\mathbf{X}}) -  \bU^{\mathsf T} \widehat{\bA} (\widehat{\mathbf{X}}- \widetilde{\mathbf{X}})\|_F  \\
	 	&\leq \sqrt{\alpha} \|\mathbf{B}-\bA\widetilde{\mathbf{X}}'\|_F+ \sqrt{\tau} + \| \bU^{\mathsf T} \bOmega^{\mathsf T} \bOmega U - \I \|_2 \cdot \| \bU^{\mathsf T} \widehat{\bA} (\widehat{\mathbf{X}}- \widetilde{\mathbf{X}}) \|_F 
\end{align*}

Using Lemma~\ref{cor:linear} and rearranging the terms, we get $\| \bU^{\mathsf T} \widehat{\bA} (\widehat{\mathbf{X}}- \widetilde{\mathbf{X}}) \|_F \leq 2\sqrt{\alpha} \| \mathbf{B} - \bA\widehat{\mathbf{X}}' \|_F + \sqrt{\tau}$. The utility proof is now immediate by observing that the column-space of $\widehat{\bA}$ and $\bU$ are the same, $\widetilde{\mathbf{X}}'$ is the optimal solution of the regression (which from the normal equations of $\widetilde{\mathbf{X}}$ implies $\bU^{\mathsf T} (\widehat{\bA} \widetilde{\mathbf{X}} - \widehat{\mathbf{B}}) = \widehat{\bA}^{\mathsf T} (\widehat{\bA} \widetilde{\mathbf{X}} - \widehat{\mathbf{B}}) =0$), and the Pythagorus theorem on the norms. More concretely,  with probability at least $1-2\beta$
\begin{align*}
	\| \bA \widehat{\mathbf{X}}' -\mathbf{B} \|^2_F &= \| A\widetilde{\mathbf{X}}' -\mathbf{B} \|^2_F + \| \bA(\widehat{\mathbf{X}}'-\widetilde{\mathbf{X}}')\|^2_F \\
				&\leq (1+4\alpha) \| \bA\widetilde{\mathbf{X}}' - \mathbf{B} \|_F +\tau.
\end{align*} 
Adjusting and renaming the values of $\alpha$ and $\beta$, we get the claim of the theorem.

\end{proof}
\end{full}
\begin{extended}
\noindent {\em Proof Sketch.} The privacy proof follows as in~\thmref{prod}. The space bound is also straightforward. The utility bound follows from taking into account the changes made to the input matrices by our mechanism and an application of the result of Kane and Nelson~\cite{KN14} to the proof idea of Sarlos~\cite{Sarlos06}. Sarlos~\cite{Sarlos06} observed that the utility guarantee for \linear \ is provided as long as $\bOmega$ provides \mult \ with multiplicative error $\sqrt{\alpha/d}$, and $\bOmega$ is an $O(d)$-space embedding. The latter follows from the standard result on any Johnson-Lindenstrauss transform with $r=O(\mathsf{rank} (A) + \log (1/\beta) /\alpha^2)$. Using~\cite{IM98}, we only need $r=O(d^2\alpha^{-1} \log(1/\beta))$ as stated in~\thmref{linear}. Note that the additive error bound is worse than~\thmref{prod} due to the difference in the value of $r$. The details of the proof is in~\appref{linear}.
\end{extended}

\medskip
\noindent{\sc \linear \ in other models.} \linear \ has been also studied in the  local privacy model by Duchi  {\it et al.}~\cite{DJW13} and in the online learning model by Jain {\it et al.}~\cite{JKT12} and Thakurata and Smith~\cite{TS13}. These models are different from ours and, therefore, our results are incomparable to theirs. However, if one wishes to make any comparison with our result, we give a brief overview of these results. The readers are welcome to compare it with~\thmref{linear}. Jain  {\it et al.}~\cite{JKT12} used  $R$ queries and $T$  training data set to give a bound of $\tilde{O} \paren{(R^6 \log (1/\delta) \sqrt{n} \log^{1.5} T )/ \sqrt{\varepsilon} \alpha^3}$. Thakurata and Smith~\cite{TS13} gave a generic algorithm of which Jain {\it et al.}~\cite{JKT12} is a special case. Duchi {\it et al.}~\cite{DJW13} gave a characterization of \linear \ in the local-privacy model.  

\subsection{Applications  in Learning Theory} Due to lack of space, we do not cover few applications and just make a short note here. For example, $\mathsf{PSG}$ can be used in manifold learning. Here, we can consider the streaming version of Baranuik and Wakin~\cite[Sec 3]{BW09}, where the sampled points are streamed. We first perform a transformation as done in \linear \ on their sampled points to boost the singular values of the matrix formed by the set of all sampled points and then use Gaussian matrix as the Johnson-Lindenstrauss transform in their algorithm. The more formal treatment of this is done by Upadhyay~\cite{Upadhyay14}. Also, our mechanism for $\lra$ can be easily compiled to give differentially private {\em principal component analysis}  using standard algorithms that use $\lra$ on the input matrix as the first step. This achieves almost the same memory bound as achieved by~\cite{MCJ13}.  


\pagebreak

\bibliographystyle{plain}
{ \bibliography{low}}

\begin{extended}
\begin{appendix}

\section{Useful Facts Used in the Paper} \label{app:facts}
\paragraph{\scshape Linear Algebra.}
Our analysis make extensive use of linear algebra and statistical properties of Gaussian distribution. We give an exposition to the level required to understand this paper.

\begin{lemma} \label{lem:conform}
	If matrix $\bA$ and $\mathbf{B}$ are conforming, then $\|\bA \mathbf{B}\|_F \leq \|\bA\|_2 \|\mathbf{B} \|_F.$
\end{lemma}

\begin{lemma} \label{lem:hermitian} 
	Let $\bA$ and $\mathbf{B}$ be Hermittian matrices. If $\bA$ and $\mathbf{B}$ differ in at most one row of Euclidean norm $1$, then $\tr(\bA^{\mathsf T}\bA) - \tr(\mathbf{B}^{\mathsf T} \mathbf{B}) \leq 2$.
\end{lemma}

\paragraph{\scshape Univariate and Multivariate Gaussian Distribution.}

\paragraph{\scshape Differential Privacy.}
We use the following in our analysis explicitly or implicitly.
\begin{theorem} {\em (\cite{DRV10}).} \label{thm:DRV}
	Let $\varepsilon, \delta \in (0,1)$, and $\delta'>0$. If $\cK_1, \cdots , \cK_\ell$ are each $(\varepsilon, \delta)$-differential private mechanism, then the mechanism $\cK(D):= (\cK_1(D), \cdots , \cK_\ell(D))$ releasing the concatenation of each algorithm is $(\varepsilon', \ell \delta+\delta')$-differentially private for $\varepsilon' < \sqrt{2\ell \ln (1/\delta')}\varepsilon + 2\ell \varepsilon^2$.
\end{theorem}
\begin{lemma} \label{lem:post}
Let $M(D)$ be a $(\varepsilon, \delta)$-differential private mechanism for a database $D$ , and let $h$ be any function, then any mechanism $M':=h(M(D))$ is also $(\varepsilon,\delta)$-differentially private for the same set of queries.
\end{lemma}

\section{Missing Proofs}
\subsection{Proof of~\thmref{PSG}(ii)} \label{app:wishart} 			 
\subsection{Proof of~\thmref{lsi}} \label{app:lra}					
\subsection{Proof of~\thmref{prod}} \label{app:prod} 		\Jcom{Rewrite the proof}		
\subsection{Utilty Proof of~\thmref{linear}} \label{app:linear} \Jcom{Rewrite the proof}			
\subsection{Proof of~\thmref{PSG}(i)} \label{app:proofs}\label{app:dp1} 



\end{appendix}
\end{extended}

\end{document}